\documentclass[sigconf]{acmart}

\usepackage{subcaption}
\usepackage{mathptmx}
\usepackage{amsmath}
\usepackage{multicol}
\usepackage{wrapfig}
\usepackage{array}
\usepackage{amssymb}
\usepackage{tikz}
\usetikzlibrary{arrows,shapes,chains,matrix,positioning,scopes,patterns}
\usepackage[linesnumbered,ruled,vlined]{algorithm2e}
\usepackage{graphicx}
\usetikzlibrary{decorations.pathreplacing}
\usepackage {enumerate}
\makeatletter
\def\@copyrightspace{\relax}
\makeatother
\renewcommand\footnotetextcopyrightpermission [1]{}

\begin{document}
\title{Content--based Dynamic Routing in Structured Overlay Networks}
\author{Muhammad Shafique}
\affiliation {Alumni Carleton University Ottawa Canada}
\email {alummi@acm.org}	

\begin{abstract}
Acyclic overlays used for broker--based publish/subscribe systems provide unique paths for content--based routing from a publisher to interested subscribers. Cyclic overlays may provide multiple paths, however, the \textit{subscription broadcast process} generates one content-based routing path per subscription. This poses serious challenges in offering \textit{dynamic routing} of notifications when congestion is detected because instantaneous updates in routing tables are required to generate alternative routing paths. This paper introduces the first subscription--based publish/subscribe system, \texttt{OctopiS}, which offers inter--cluster dynamic routing when congestion in the output queues is detected. \texttt{OctopiS} is based on a formally defined \textit{Structured Cyclic Overlay Topology (SCOT)}. SCOT is divided into \textit{homogeneous clusters} where each cluster has equal number of brokers and connects to other clusters through multiple \textit{inter--cluster} overlay links. These links are used to provide parallel routing paths between publishers and subscribers connected to brokers in different clusters. While aiming at deployment at data center networks, \texttt{OctopiS} generates subscription--trees of shortest lengths used by \textit{Static Notification Routing (SNR)} algorithm. \textit{Dynamic Notification Routing (DNR)} algorithm uses a bit--vector mechanism to exploit the \textit{structuredness} of a clustered SCOT to offer inter--cluster dynamic routing without making updates in routing tables and minimizing load on overwhelmed brokers and congested links. Experiments on a cluster testbed with real world data show that \texttt{OctopiS} is scalable and reduces the number of inter--broker messages in subscription delivery by 89\%, subscription delay by 77\%, end--to--end notification delay in static and dynamic routing by 47\% and 58\% respectively, and the lengths of output queues of brokers in dynamic routing paths by 59\%.
\end{abstract}

\maketitle
\raggedbottom

\section{Introduction}
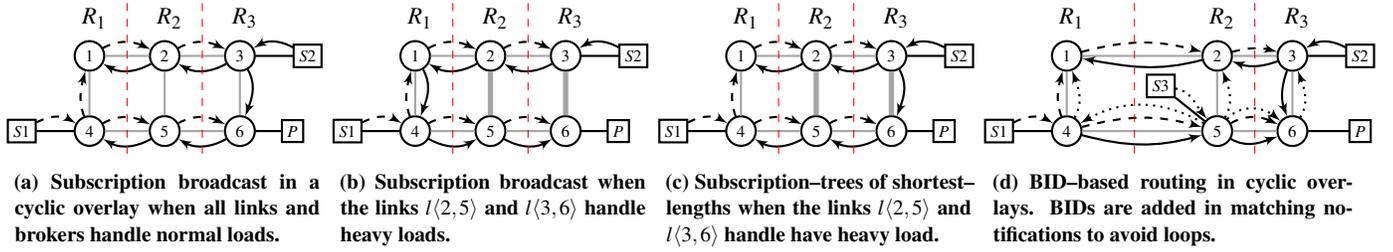
\begin{figure*}
	%define line/arrow styles...
	\tikzstyle{line} = [draw, -latex']
	\def\scaleFac {0.7}
	\begin{subfigure}[b]{0.24\textwidth}
		\captionsetup{width=0.95\linewidth}
		\begin{tikzpicture}
		\captionsetup{width=10cm}
		\def\xInc {1}
		\def\x {0}
		\def\y {0}
		\def\yInc {1}
		\tikzstyle{every node} = [thick, scale=1]
		\node[draw, circle, scale=\scaleFac] (1) at (\x+\xInc,\y+\yInc) 			{$1$};
		\node[draw, circle, scale=\scaleFac] (2) at (\x+\xInc*2,\y+\yInc) 		{$2$};
		\node[draw, circle, scale=\scaleFac] (3) at (\x+\xInc*3,\y+\yInc) 		{$3$};
		\node[draw, circle, scale=\scaleFac] (4) at (\x+\xInc,\y) 			{$4$};
		\node[draw, circle, scale=\scaleFac] (5) at (\x+\xInc*2,\y) 			{$5$};
		\node[draw, circle, scale=\scaleFac] (6) at (\x+\xInc*3,\y) 			{$6$};
		%define the clients here...
		\node[draw, rectangle, scale=0.7] (S1)	at (\x+0.1,0) 					{$S1$};
		\node[draw, rectangle, scale=0.7] (P)		 at (\x+\xInc*4-0.3, 0)		{$P$};
		\node[draw, rectangle, scale=0.7] (S2)	at (\x+\xInc*4-0.1,\y+\yInc) 	 	{$S2$};
		\draw [color=gray!70, thick]  (5) -- (4) (6) -- (5) (3) -- (6) (2) -- (3) (1) -- (2) (4) -- (1) (5) -- (2); 
		\draw [thick] (S1) -- (4) (P) -- (6) (S2) -- (3);	
		%draw arrows for S1
		\path [line, dashed, thick] (S1) to [out=30,in=150] (4);
		\path [line, dashed, thick] (4) to [out=30,in=155] (5);
		\path [line, dashed, thick] (4) to [out=110,in=255] (1);
		\path [line, dashed, thick] (5) to [out=30,in=155] (6);
		\path [line, dashed, thick] (1) to [out=30,in=155] (2);
		\path [line, dashed, thick] (2) to [out=30,in=155] (3);
		%draw arrows for S2
		\path [line, thick] (S2) to [out=150,in=30] (3);
		\path [line, thick] (3) to [out=210,in=330] (2);
		\path [line, thick] (2) to [out=210,in=330] (1);
		\path [line, thick] (6) to [out=210,in=330] (5);
		\path [line, thick] (5) to [out=210,in=330] (4);
		\path [line, thick] (3) to [out=300,in=70] (6);
		%draw the lines to indicate regions...
		\draw[color=red, dashed] (\x+1.5,-0.3) -- (\x+1.5,1.8);
		\draw[color=red, dashed] (\x+2.5,-0.3) -- (\x+2.5,1.8);		
		%draw the region text
		\node[text width=3cm] at (\x+2.4, 1.5) {$R_1$};
		\node[text width=3cm] at (\x+3.4,1.5) {$R_2$};
		\node[text width=3cm] at (\x+4.4,1.5) {$R_3$};
		\end{tikzpicture}
		\caption{Subscription broadcast in a cyclic overlay when all links and brokers handle normal loads.}
		\label{fig:BK1}		
	\end{subfigure}
	~ %add desired spacing between images, e. g. ~, \quad, \qquad, \hfill etc.
	\begin{subfigure}[b]{0.24\textwidth}
		\captionsetup{width=0.95\linewidth}
		\begin{tikzpicture}
		\def\xInc {1}
		\def\x {0}
		\def\y {0}
		\def\yInc {1}
		\tikzstyle{every node} = [thick, scale=1]
		\node[draw, circle, scale=\scaleFac] (1) at (\x+\xInc,\y+\yInc) 			{$1$};
		\node[draw, circle, scale=\scaleFac] (2) at (\x+\xInc*2,\y+\yInc) 		{$2$};
		\node[draw, circle, scale=\scaleFac] (3) at (\x+\xInc*3,\y+\yInc) 		{$3$};
		\node[draw, circle, scale=\scaleFac] (4) at (\x+\xInc,\y) 			{$4$};
		\node[draw, circle, scale=\scaleFac] (5) at (\x+\xInc*2,\y) 			{$5$};
		\node[draw, circle, scale=\scaleFac] (6) at (\x+\xInc*3,\y) 			{$6$};
		%define the clients here...
		\node[draw, rectangle, scale=0.7] (S1)	at (\x+0.1,0) 					{$S1$};
		\node[draw, rectangle, scale=0.7] (P)		 at (\x+\xInc*4-0.3,0)	{$P$};
		\node[draw, rectangle, scale=0.7] (S2)	at (\x+\xInc*4-0.1,\y+\yInc) 	 	{$S2$};
		\draw [color=gray!70, thick]  (5) -- (4) (6) -- (5) (2) -- (3) (1) -- (2) (4) -- (1); 
		\draw [thick] (S1) -- (4) (P) -- (6) (S2) -- (3);
		\draw[line width=2, color=gray!70] (5) -- (2) (3) -- (6);
		%draw arrows for S1
		\path [line, dashed, thick] (S1) to [out=30,in=150] (4);
		\path [line, dashed, thick] (4) to [out=30,in=155] (5);
		\path [line, dashed, thick] (4) to [out=110,in=255] (1);
		\path [line, dashed, thick] (5) to [out=30,in=155] (6);
		\path [line, dashed, thick] (1) to [out=30,in=155] (2);
		\path [line, dashed, thick] (2) to [out=30,in=155] (3);
		%draw arrows for S2
		\path [line, thick] (S2) to [out=150,in=30] (3);
		\path [line, thick] (3) to [out=210,in=330] (2);
		\path [line, thick] (2) to [out=210,in=330] (1);
		\path [line, thick] (5) to [out=330,in=210] (6);
		\path [line, thick] (4) to [out=330,in=210] (5);
		\path [line, thick] (1) to [out=300,in=70] (4);
		%draw the lines to indicate regions...
		\draw[color=red, dashed] (\x+1.5,-0.3) -- (\x+1.5,1.8);
		\draw[color=red, dashed] (\x+2.5,-0.3) -- (\x+2.5,1.8);		
		%draw the region text
		\node[text width=3cm] at (\x+2.4, 1.5) {$R_1$};
		\node[text width=3cm] at (\x+3.4,1.5) {$R_2$};
		\node[text width=3cm] at (\x+4.4,1.5) {$R_3$};
		\end{tikzpicture}
		\caption{Subscription broadcast when the links $l \langle 2,5\rangle$ and $l \langle 3,6\rangle$ handle heavy loads.}
		\label{fig:BK2}
	\end{subfigure}
	~
	\begin{subfigure}[b]{0.24\textwidth}
		\captionsetup{width=0.95\linewidth}
		\begin{tikzpicture}
		\def\xInc {1}
		\def\x {0}
		\def\y {0}
		\def\yInc {1}
		\tikzstyle{every node} = [thick, scale=1]
		\node[draw, circle, scale=\scaleFac] (1) at (\x+\xInc,\y+\yInc) 			{$1$};
		\node[draw, circle, scale=\scaleFac] (2) at (\x+\xInc*2,\y+\yInc) 		{$2$};
		\node[draw, circle, scale=\scaleFac] (3) at (\x+\xInc*3,\y+\yInc) 		{$3$};
		\node[draw, circle, scale=\scaleFac] (4) at (\x+\xInc,\y) 			{$4$};
		\node[draw, circle, scale=\scaleFac] (5) at (\x+\xInc*2,\y) 			{$5$};
		\node[draw, circle, scale=\scaleFac] (6) at (\x+\xInc*3,\y) 			{$6$};
		%define the clients here...
		\node[draw, rectangle, scale=0.7] (S1)	at (\x+0.1,0) 					{$S1$};
		\node[draw, rectangle, scale=0.7] (P)		 at (\x+\xInc*4-0.3,0)	{$P$};
		\node[draw, rectangle, scale=0.7] (S2)	at (\x+\xInc*4-0.1,\y+\yInc) 	 	{$S2$};
		\draw [color=gray!70, thick]  (5) -- (4) (6) -- (5) (2) -- (3) (1) -- (2) (4) -- (1); 
		\draw [thick] (S1) -- (4) (P) -- (6) (S2) -- (3);
		\draw[line width=2, color=gray!70] (5) -- (2) (3) -- (6);
		%draw arrows for S1
		\path [line, dashed, thick] (S1) to [out=30,in=150] (4);
		\path [line, dashed, thick] (4) to [out=30,in=155] (5);
		\path [line, dashed, thick] (4) to [out=110,in=255] (1);
		\path [line, dashed, thick] (5) to [out=30,in=155] (6);
		\path [line, dashed, thick] (1) to [out=30,in=155] (2);
		\path [line, dashed, thick] (2) to [out=30,in=155] (3);
		%draw arrows for S2
		\path [line, thick] (S2) to [out=150,in=30] (3);
		\path [line, thick] (3) to [out=210,in=330] (2);
		\path [line, thick] (2) to [out=210,in=330] (1);
		\path [line, thick] (6) to [out=210,in=330] (5);
		\path [line, thick] (5) to [out=210,in=330] (4);
		\path [line, thick] (3) to [out=300,in=70] (6);
		%draw the lines to indicate regions...
		\draw[color=red, dashed] (\x+1.5,-0.3) -- (\x+1.5,1.8);
		\draw[color=red, dashed] (\x+2.5,-0.3) -- (\x+2.5,1.8);		
		%draw the region text
		\node[text width=3cm] at (\x+2.4, 1.5) {$R_1$};
		\node[text width=3cm] at (\x+3.4,1.5) {$R_2$};
		\node[text width=3cm] at (\x+4.4,1.5) {$R_3$};				
		\end{tikzpicture}
		\caption{Subscription--trees of shortest--lengths when the links $l \langle 2,5\rangle$ and $l \langle 3,6\rangle$ handle have heavy load.}
		\label{fig:BK3}
	\end{subfigure}
	~
	\begin{subfigure}[b]{0.28\textwidth}
		\captionsetup{width=0.95\linewidth}
		\begin{tikzpicture}
		\def\xInc {1.5}
		\def\x {0}
		\def\y {0}
		\def\yInc {1}
		\tikzstyle{every node} = [thick, scale=1]
		\node[draw, circle, scale=\scaleFac] (1) at (\x+\xInc-0.5,\y+\yInc)		{$1$};
		\node[draw, circle, scale=\scaleFac] (2) at (\x+\xInc*2,\y+\yInc) 		{$2$};
		\node[draw, circle, scale=\scaleFac] (3) at (\x+\xInc*3-0.5,\y+\yInc)	{$3$};
		\node[draw, circle, scale=\scaleFac] (4) at (\x+\xInc-0.5,\y) 			{$4$};
		\node[draw, circle, scale=\scaleFac] (5) at (\x+\xInc*2,+\y) 			{$5$};
		\node[draw, circle, scale=\scaleFac] (6) at (\x+\xInc*3-0.5,+\y) 		{$6$};
		%define the clients here...
		\node[draw, rectangle, scale=0.7] (S1)	at (\x+0.1,\y) 						{$S1$};
		\node[draw, rectangle, scale=0.7] (P)		 at (\x+\xInc*4-1.1,\y)		{$P$};
		\node[draw, rectangle, scale=0.7] (S2)	at (\x+\xInc*4-1.1,\y+\yInc)	{$S2$};
		\node[draw, rectangle, scale=0.7] (S3)	at (\x+\xInc*1.5,\y+0.6) 	 	{$S3$};
		\draw [color=gray!70, thick]  (5) -- (4) (6) -- (5) (2) -- (3) (1) -- (2) (4) -- (1); 
		\draw [thick] (S1) -- (4) (P) -- (6) (S2) -- (3) (S3)--(5);
		\draw[thick, color=gray!70] (5) -- (2) (3) -- (6);
		%draw arrows for S1
		\path [line, dashed, thick] (S1) to [out=30,in=150] (4);
		\path [line, dashed, thick] (4) to [out=15,in=165] (5);
		\path [line, dashed, thick] (4) to [out=110,in=255] (1);
		\path [line, dashed, thick] (5) to [out=30,in=155] (6);
		\path [line, dashed, thick] (1) to [out=15,in=165] (2);
		\path [line, dashed, thick] (2) to [out=15,in=165] (3);
		%draw arrows for S2
		\path [line, thick] (S2) to [out=150,in=30] (3);
		\path [line, thick] (3) to [out=200,in=340] (2);
		\path [line, thick] (2) to [out=195,in=350] (1);
		\path [line, thick] (5) to [out=330,in=200] (6);
		\path [line, thick] (4) to [out=345,in=195] (5);
		\path [line, thick] (3) to [out=250,in=110] (6);
		%draw arrows for S3
		\path [line, thick, dotted] (S3) to [out=350,in=130] (5);
		\path [line, thick, dotted] (5) to [out=60,in=290] (2);
		\path [line, thick, dotted] (6) to [out=60,in=290] (3);
		\path [line, thick, dotted] (4) to [out=60,in=290] (1);
		\path [line, thick, dotted] (5) to [out=50,in=140] (6);
		\path [line, thick, dotted] (5) to [out=150,in=30] (4);
		%draw the lines to indicate regions...
		\draw[color=red, dashed] (\x+1.9,-0.3) -- (\x+1.9,1.8);
		\draw[color=red, dashed] (\x+3.5,-0.3) -- (\x+3.5,1.8);		
		%draw the region text
		\node[text width=3cm] at (\x+2.4,1.5) {$R_1$};
		\node[text width=3cm] at (\x+4.4,1.5) {$R_2$};
		\node[text width=3cm] at (\x+5.4,1.5) {$R_3$};				
		\end{tikzpicture}
		\caption{BID--based routing in cyclic overlays. BIDs are added in matching notifications to avoid loops.}
		\label{fig:BK4}
	\end{subfigure}
	\caption{A cyclic overlay of six brokers each represented by a circle. The grey lines indicate overlay links that connect brokers. The thick grey lines are overloaded links. The dashed, solid and dotted arrow messages indicate the subscription--trees of the subscribers S1, S2, and S3 respectively, interested in notifications from the publisher P. The dashed red lines separate regions $R_{1}$, $R_{2}$, and $R_{3}$.} 
	\label{fig:Subscription}
\end{figure*}
\textit{Content--based Publish/Subscribe (CPS)} systems are used for many--to--many communication among loosely coupled distributed entities. A \textit{publisher} publishes data in form of \textit{notifications}, while a \textit{subscriber} registers its interest in form of a \textit{subscription} (set of filters or predicates) to receive notifications of interest \cite{MANY_FACES,carz_thesis,PADRESBookChapte,Tarkoma}. In \textit{broker--based} CPS systems, a dedicated \textit{overlay network}, formed by a set of inter--connected brokers, is used to connect publishers and subscribers while keeping them anonymous from each other \cite{MANY_FACES,carz_thesis}.

Publish/subscribe is an active area of research due to its increasing popularity and gradual adoption in different application domains \cite{Tarkoma}. It is the communication substratum in social networking systems \cite{KaiwenZhang_1,WormHole}, business process monitoring \cite{muthy_thesis}, software defined networking \cite{TariqPLEROMA}, massive multi--player online games \cite{MMOG_Canas}, and many commercial applications \cite{PNUTS, G_CLOUD_PS, MS_PULSE, KAFKA,WormHole}. A subscription is broadcast in the overlay network and saved in \textit{routing table} of each broker in order to form a subscription--tree. Upon receiving a notification, a broker calculates next destination--paths by matching contents in the notification with filters in saved subscriptions. This technique is called the \textit{content--based} or \textit{filter--based} routing using \textit{Reverse Path Forwarding} \cite{RPF, carz_thesis}. A CPS system in this paper refers to a subscription--based publish/subscribe system that uses a dedicated broker--based overlay network for content--based routing \cite{SIENA_WIDE_AREA}. 

Most CPS systems use acyclic overlay topologies, which provide single routing path and offers limited flexibility to deal with network conditions like load imbalance, and link congestion. To stabilize a CPS system, subscribers are shifted from overloaded to less loaded brokers in a network area. Unfortunately, finding less loaded brokers requires extra in-broker processing and generates additional network traffic, which exerts more load on the system. This not only keeps a system unstable for quite sometime until the load shift process is complete but also causes loss of messages \cite{dynamic_LoadBalancing}. Cyclic overlay networks are expected to improve performance and throughput by offering multiple paths among publishers and subscribers. When a link congestion or load imbalance is detected, the best available path can be selected to offer \textit{dynamic routing}. Although multiple paths may be available, at most one routing path, activated by a \textit{subscription--tree}, can be used to route notifications \cite{SIENA_WIDE_AREA}. If a link is congested, no alternative routing paths are available for content--based routing. The available multiple links can be exploited to generate new content-based routing paths and avoid shifting subscribers, however, this requires an intelligent algorithm to search for alternative routing paths and then make a large number of updates in routing tables, which is costly and not scalable for large network settings (cf. Sec. 2). This indicates that for dynamic routing, acyclic and cyclic overlays suffer from almost the same limitations. Ideally, for high throughput and scalability, dynamic routing should be achieved without requiring updates in routing tables. Notification routing in cyclic overlay networks has received a little attention and, to the best of our knowledge, no CPS system offers dynamic routing. In addition to the limitation of one subscription--tree, the traditional CPS systems have more fundamental issues. For example, each subscription should be uniquely identified to avoid loops. Notifications should carry identifications of matching subscriptions to identify routing paths. Extra inter-broker messages, and larger lengths of routing paths with no support for dynamic routing (cf. Sec. 2).  

This paper introduces the first CPS system, \texttt{OctopiS}, which offers \textit{inter--cluster} dynamic routing. The system is based on a purpose-built topology called \textit{Structured Cyclic Overlay Topology (SCOT)} generated from Cartesian product of graphs (cf. Sec. 3). We use a novel \textit{clustering} technique to divide SCOT into groups of brokers (i.e., \textit{clusters}). Classifications of clusters, brokers, and links are introduced to define \text{structuredness} of a SCOT (cf. Sec. 4). \texttt{OctopiS} exploits the \text{structuredness} to generate subscription--trees of shortest lengths and do not require unique identifications to detect loops. The system offers \textit{inter-cluster} dynamic routing without making updates in routing tables (cf. Secs. 4 \& 5). The proposed \textit{Static Notification Routing (SNR)} algorithm uses subscription--trees to send notifications to interested subscribers, while \textit{Dynamic Notification Routing (DNR)} algorithm reduces delivery delay by offering inter--cluster dynamic routing when congestion is detected. In summary, the contributions of this paper are as follows. (i) Sec. 2 identifies issues surrounding the CPS systems that use cyclic overlays. (ii) Sec. 3 introduces how Cartesian product of graphs can be used to formally describe SCOT. Additional constraints are introduced to optimize SCOT for content--based routing. (iii) Sec. 4 describes a clustering approach with additional classifications for topology elements to prevent loops in SCOT. This section also introduces a lightweight bit--vector mechanism to identify \textit{target clusters} for inter--cluster dynamic routing. (iv) A subscription broadcast algorithm that generates subscription--trees of shortest lengths in a clustered SCOT is described in Sec. 5. (v) Details on static routing (by \textit{SNR} algorithm) and dynamic routing (by \textit{DNR} algorithm) are provided in Sec. 6. (vi) Comparison with state--of--the--art \textit{identification--based} routing is discussed in Sec. 7. We describe related work in Sec. 8, and conclude in Sec. 9.
\section{Background Issues}
In this section, we use Fig. 1 to discuss different issues related to content--based routing in cyclic overlays. In particular, we discuss the issue of: (i) adding a \textit{Unique Identification} to each subscription to avoid cycles in subscription broadcast, (ii) \textit{Extra Inter--broker Messages (IMs)} to detect and discard duplicate subscriptions, (iii) larger \textit{Lengths of Subscription--trees}, and (iv) \textit{Path Identification} for notification routing. In this paper, an overlay link is represented as $l\langle source, destination\rangle$, where the \textit{source} and \textit{destination} are message sending and receiving brokers respectively.\\
\textbf{I1 Unique Identification}: Content--based routing generates loops in cyclic overlays. Loops route a message indefinitely if it is not detected and discarded. The \textit{Subscription Broadcast Process (SBP)} broadcasts subscriptions to form subscription--trees for notifications routing. Since multiple paths can be available, a broker may receive duplicate subscriptions (or duplicates). To solve this issue, each broker adds its unique identification, called \textit{Broker Identification (or BID)}, in a subscription of the \textit{local subscriber}. The subscription and BID form a network-wide unique identification, which is saved in routing tables to detect and discard duplicates.\\ 
\textbf{I2 Extra Inter--broker Messages (IMs)}: Extra IMs have to be generated to detect and discard duplicates. In Fig. 1(a), broker 4 receives the subscription of S2 from broker 5. The link $l \langle 5,4\rangle$ is added in the subscription--tree of S2, assuming that broker 4 discards a second copy of the subscription received from broker 1. Similarly, broker 5 discards a second copy of the subscription of S1 received from broker 2, assuming that the first copy was already received from broker 4 (there is, therefore, no subscription--tree link from broker 2 to broker 5). This indicates that, despite using BIDs, extra IMs are generated to detect and discard duplicates.\\ 
\textbf{I3 Subscription--tree Length}: In an acyclic overlay, SBP generates a unique subscription--tree even if a subscription is issued multiple times (after calling unsubscribe from the same broker). However, this is not the case in cyclic overlays, where multiple subscribe calls issued from the same broker may generate multiple subscription--trees with different lengths. SBP in cyclic overlays is an \textit{uncontrolled} process and selects the first available link (or broker) as the next destination. This may generate subscription--trees of larger lengths when load on the links and brokers is uneven. For example, subscription--trees of S1 and S2 in Fig. 1(a) have shortest lengths (number of hops), however, the subscription--tree of S2 in Fig. 1(b) has larger length. Presumably, the links $l \langle 2,5\rangle$ and $l \langle 3,6\rangle$ had heavy network traffic when S2 issued the subscription. Broker 6 received the subscription of S2 from broker 5 and discarded the duplicate received from broker 3. Although P and S2 are hosted by the brokers in the same region (i.e., $R_{3}$), S2 receives notifications from P after they are processed by brokers in $R_{1}$ and $R_{2}$. Subscription--trees of longer lengths increase in--broker computation, generate extra IMs, waste network bandwidth, and cause high latency in notification delivery. Ideally, a subscription--tree should always has the shortest length, even if some links have high loads when the subscription is issued (e.g., the subscription--trees in Fig. 1(c)). To the best of our knowledge, no CPS system generates subscription--trees of shortest lengths.\\
\textbf{I4 Path Identification}: The issue of loops is also relevant for delivering notifications. For example, the matching process executed at broker 6 in Fig. 1(a) indicates that a notification $n$ from P should be forwarded onto the links $l \langle6,3\rangle$ and $l \langle6,5\rangle$. Broker 6 creates two copies of $n$, $n_{1}$ and $n_{2}$, to forward to brokers 5 and 3, respectively. S1 receives $n_{1}$ from broker 4 and S2 receives $n_{2}$ from broker 3. Unfortunately, the matching process at broker 3 indicates that the subscription of S1 matches $n_{2}$, and a copy of $n_{2}$, say $n_{2}^{'}$, should be forwarded to broker 2. $n_{2}^{'}$ ultimately reaches S1 after being forwarded by broker 4. Again, the matching process at broker 4 identifies that S2 should also receive a copy of $n_{2}^{'}$, and this process continues indefinitely until identified and stopped. Similarly, $n_{1}$, forwarded by broker 6 onto $l \langle6,5\rangle$, is received more than once. To prevent receiving duplicates, host broker of a publisher adds BIDs, assigned to matching subscriptions, to a notification. Reverse path forwarding technique is used to route the notification along the paths identified by the BIDs and each interested broker removes its BID from the notification before forwarding it to the next brokers. Routing is stopped when no BID is left in the notification. This approach is further explained by using Fig. 1(d). For any notification $m$ from P, broker 6 creates two copies, $m_{1}$ and $m_{2}$, where $m_{1}$ with the BID of broker 3 is forwarded onto link $l \langle6,3\rangle$, and $m_{2}$ with the BIDs of brokers 4 and 5 is forwarded onto link $l \langle6,5\rangle$. Because of carrying an additional BID, payload of $m_{2}$ is greater than payload of $m_{1}$. After receiving $m_{2}$, broker 5 removes its own BID and forwards one copy of $m_{2}$ to S3, and another copy to broker 4. After receiving $m_{2}$, broker 4 removes its own BID and forwards a copy of $m_{2}$ to S1. Since no BID is left, broker 4 does not forward $m_{2}$ any further. In a large overlay network, a notification may be received by a large number of subscribers hosted by many brokers and many BIDs may have to be added in a notification \cite{MS_PULSE}. In scenarios where scalability is a major requirement, BID--based routing is a bottleneck.\\ 
\textbf{I5 Single Routing Path}: As SBP generates one subscription--tree per subscription, updates in routing tables have to made to offer dynamic routing. Fig. 1(a) shows that there are three paths from broker 6 to broker 3 and only the path (with link $l \langle3, 6\rangle$), being part of the subscription--tree of S2, is used as a content-based routing path. If the link $l \langle6,3\rangle$ is congested, S2 has to be moved to some other broker in the less loaded network area, which requires \textit{unsubscribe} and \textit{subscribe} calls generating more network traffic and requiring updates in a number of routing tables. If the link $l \langle6,3\rangle$ is broken for some reason, a new subscription--tree has to be generated to send notifications to S2. This requires an intelligent algorithm that makes updates in routing tables of brokers 2, 3, 5, and 6 to remove the link $l \langle3, 6\rangle$ and add the link $l \langle2, 5\rangle$. Looking at the decoupled nature of CPS systems, such algorithms are difficult to design and not scalable for large networks.  
\section{Structured Cyclic Topology}
In this section we describe our approach of designing a structured cyclic topology for loop free content-based routing. Structured cyclic topologies provide parallel links that can be used as alternative routing paths when congestion is detected. Our goal is to use inter-cluster parallel links as alternative content-based routing paths when congestion is detected. This requires no updates in routing tables to offer inter-cluster dynamic routing. We use the Cartesian Product of Undirected Graphs (CPUG) to design large, structured overlay cyclic networks based on small graph patterns \cite{theGeneralizedPUG}.
\subsection{Preliminaries}
A graph is an ordered pair $G = (V_{G}, E_{G})$, where $V_{G}$ is a finite set of vertices and $E_{G}$ is a set of edges or links that connect two vertices in $G$. The number of vertices of $G$ (called \textit{order}) is $\mid G\mid$ (or $|V_{G}|$). Similarly, the number of edges in $G$ is $\parallel G \parallel$ (or $|E_{G}|$). A graph in which each pair of vertices are connected by an edge is called a \textit{complete graph}. The diameter of a graph G, represented as \textit{diam(G)}, is the shortest path between the two most distant nodes in $G$.

A graph product is a binary operation that takes two small graph operands---for example $G(V_{G}, E_{G})$ and $H(V_{H}, E_{H})$---to produces a large graph whose vertex set is given by $V_{G} \mathsf{X} V_{H}$. Many types of graph products exist, but we find the Cartesian product most suitable for content-based routing. Other products, for example, the Direct product and the Strong product can be used but their rule-based interconnection of vertices increases node degree and makes routing complex. The $CPUG$ of two graphs $G(V_{G}, E_{G})$ and $H(V_{H}, E_{H})$ is denoted by $G  \square  H$, with vertex set $V_{G \Box H}$ and set of edges $E_{G \Box H}$. Two vertices $(g, h) \in V_{G \Box H}$ and $(g', h') \in V_{G \Box H}$ are adjacent if $g=g'$ and $hh' \in E_{G \Box H}$ or $gg' \in E_{G \Box H}$ and $h=h'$. Formally, the sets of vertices and edges of a CPUG are given as \cite{CPUG_Book}.
\begin{equation}
V_{G \square H} = \{ (g, h) | g \in V_{G}  \wedge  h \in V_{H}\}
\end{equation}
\begin{equation}
\left.\begin{aligned}
E_{G \square H} = \{ \langle (g, h)(g', h') \rangle | (g=g', hh' \in E_{H}) \\ \vee (gg' \in E_{G}, h=h')\}
\end{aligned}
\right\}
\end{equation}
The operand graphs $G$ and $H$ are called factors of $G  \square  H$. CPUG is commutative---that is, $G  \Box  H = H  \square  G$. Although CPUG of $n$ number of graphs is possible, we are concerned with CPUG of only two graphs.
\subsection{Structured Cyclic Overlay Topology}
The \textit{Structured Cyclic Overlay Topology (SCOT)} is a $CPUG$ of two graphs. One graph, represented by $G_{af}$, is called \textit{SCOT acyclic factor}, while the second graph operand, represented by $G_{cf}$, is called \textit{SCOT connectivity factor}. A SCOT has two important properties: (i) \textit{Acyclic Property} emphasizes that the $G_{af}$ must be an acyclic graph, and (ii) \textit{Connectivity Property} requires that $G_{cf}$ must be a complete graph. These properties augment a SCOT with essential characteristics that are used for generating subscription--trees of shortest lengths. $V_{af}$ and $V_{cf}$ are the sets of vertices of $G_{af}$ and $G_{cf}$, while $E_{af}$ and $E_{cf}$ are the sets of edges of $G_{af}$ and $G_{cf}$, respectively. For a \textit{singleton graph} of vertex set $\{h\} \subset V_{cf}$, the graph $G_{af}^h$ generated by $G_{af} \Box \{h\}$ is called a $G^h_{af}-fiber$ with \textit{index h}. Similarly, for a singleton graph of vertex set $\{m\} \subset G_{af}$, the graph $G^m_{cf}$ generated by $\{m\} \square G_{cf}$ is called a $G^m_{cf}-fiber$ with \textit{index m}. We describe the importance of using indexes in SCOT fibers in Section 4. The definitions of the fibers indicate that, for each vertex of $G_{cf}$, $CPUG$ generates one replica of $G_{af}$, and for each vertex of $G_{af}$, \textit{CPUG} generates one replica of $G_{cf}$. The number of distinct fibers of $G_{af}$ and $G_{cf}$ is equal to $|V_{cf}|$ and $|V_{af}|$ respectively. 
\begin{tikzpicture}
\tikzstyle{every node} = [minimum size=7mm]
\def\y {6}
\def\x {1}
\def\xInc {1}
\def\scaleFac {0.7}
\node[draw, thick, circle, scale=\scaleFac] (a) at (\x,\y) {$a$};
\node[draw, thick, circle, scale=\scaleFac] (b) at (\x+\xInc,\y) {$b$};
\node[draw, thick, circle, scale=\scaleFac] (c) at (\x+\xInc*2,\y) {$c$};
\node[draw, thick, circle, scale=\scaleFac] (d) at (\x+\xInc*3,\y) {$d$};
\node[draw, thick, circle, scale=\scaleFac] (e) at (\x+\xInc*4,\y) {$e$};
\node[draw, thick, circle, scale=\scaleFac] (f) at (\x+\xInc*5,\y) {$f$};
%draw the curved line...
\draw [thick] (b) to [out=30,in=150] (e);
\draw [thick] (a) -- (b) (b) -- (c) (d) -- (e) (e) -- (f);
%draw the operator box.
\node[draw, thick, rectangle, scale=0.4] (op)	at (\x+\xInc*6,\y) {};
%draw the connectivity triangle...
\node[draw, thick, circle, scale=\scaleFac, fill=green!50] (va) at (\x+\xInc*7,\y) {$1$};
\node[draw, thick, circle, scale=\scaleFac] (vb) at (\x+\xInc*8,\y) {$2$};
\node[draw, thick, circle, scale=\scaleFac, pattern=north west lines, pattern color=gray!40] (vc) at (\x+\xInc*7+0.5,\y+0.6) {$0$};		
%draw tirangle edges...
\draw [dashed, thick] (va) -- (vb) (vb) -- (vc) (vc) -- (va);
\end{tikzpicture}

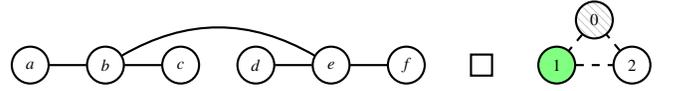
\captionof{figure}{Operands of CPUG: Left of $\Box$ is $G_{af}$ which is an H-graph; right of $\Box$ is $G_{cf}$ which is a triangle.\\}
\label{fig:Operands}
In addition to acyclic and connectivity properties, a SCOT has two more properties: (i) \textit{Index property}, which emphasizes that the labels of nodes of $G_{cf}$ must be a sequence of unique integers starting from zero, and (ii) \textit{Label Order property}, which requires that the first operand (from left to right) of a $CPUG$ node should be from node of $G_{af}$. The index property implies that the index of each fiber of  $G_{af}$ is always an integer. The label order property indicates that the first part of the label of a SCOT node comes from the corresponding vertex of $V_{af}$, and the second part is the label of the corresponding vertex of $V_{cf}$, as indicated by Eq. 1. Reversing the order of operands does not generate extra links or nodes, since CPUG is commutative. These two properties are used for clustering and routing purposes (cf. Secs. 4--6). In Fig. 2, the left operand of $\square$ operator, an acyclic $H-graph$, is the $G_{af}$, while the second operand $G_{cf}$ is a triangle, which is a complete graph. More details on SCOT are available in the technical report \cite{OctopiA_TR}.
\begin{tikzpicture}
%\footnotesize
\def\scaleSCOT {0.6}
\def\xInc {1.4}
\def\x {0}
\node[draw, thick, circle, scale=\scaleSCOT, pattern=north west lines, pattern color=gray!40] (1) at (\x,6) {$a,0$};
\node[draw, thick, circle, scale=\scaleSCOT, pattern=north west lines, pattern color=gray!40] (2) at (\x+\xInc,6) {$b,0$};
\node[draw, thick, circle, scale=\scaleSCOT, pattern=north west lines, pattern color=gray!40] (3) at (\x+\xInc*2,6) {$c,0$};
\node[draw, thick, circle, scale=\scaleSCOT, pattern=north west lines, pattern color=gray!40] (4) at (\x+\xInc*3,6) {$d,0$};
\node[draw, thick, circle, scale=\scaleSCOT, pattern=north west lines, pattern color=gray!40] (5) at (\x+\xInc*4,6) {$e,0$};
\node[draw, thick, circle, scale=\scaleSCOT, pattern=north west lines, pattern color=gray!40] (6) at (\x+\xInc*5,6) {$f,0$};
%draw the curved line...
\draw [thick] (2) to [out=15,in=165] (5);
\node[draw, thick, circle, scale=\scaleSCOT, fill=green!50] (11) at (\x,5.1) {$a,1$};
\node[draw, thick, circle, scale=\scaleSCOT, fill=green!50] (12) at (\x+\xInc,5.1) {$b,1$};
\node[draw, thick, circle, scale=\scaleSCOT, fill=green!50] (13) at (\x+\xInc*2,5.1) {$c,1$};
\node[draw, thick, circle, scale=\scaleSCOT, fill=green!50] (14) at (\x+\xInc*3,5.1) {$d,1$};
\node[draw, thick, circle, scale=\scaleSCOT, fill=green!50] (15) at (\x+\xInc*4,5.1) {$e,1$};
\node[draw, thick, circle, scale=\scaleSCOT, fill=green!50] (16) at (\x+\xInc*5,5.1) {$f,1$};
%draw the curved line...
\draw [thick] (12) to [out=15,in=165] (15);
\node[draw, thick, circle, scale=\scaleSCOT] (21) at (\x,4.2) {$a,2$};
\node[draw, thick, circle, scale=\scaleSCOT] (22) at (\x+\xInc,4.2) {$b,2$};
\node[draw, thick, circle, scale=\scaleSCOT] (23) at (\x+\xInc*2,4.2) {$c,2$};
\node[draw, thick, circle, scale=\scaleSCOT] (24) at (\x+\xInc*3,4.2) {$d,2$};
\node[draw, thick, circle, scale=\scaleSCOT] (25) at (\x+\xInc*4,4.2) {$e,2$};
\node[draw, thick, circle, scale=\scaleSCOT] (26) at (\x+\xInc*5,4.2) {$f,2$};
%draw the curved line...
\draw [thick] (22) to [out=15,in=165] (25);
\draw [thick] (1) -- (2) (2) -- (3) (4) -- (5) (5) -- (6) (11) -- (12) (12) -- (13) (14) -- (15) (15) -- (16) (21) -- (22) (22) -- (23) (24) -- (25) (25) -- (26);
%draw the dashed i-COL lines...					
\draw  [dashed, thick] (1) -- (11) (11) -- (21) (2) -- (12) (12) -- (22) (3) -- (13) (13) -- (23) (4) -- (14) (14) -- (24) 
(5) -- (15) (15) -- (25) (6) -- (16) (16) -- (26);
%drwa the curved dotted links...		
\draw [dashed, thick] (1) to [out=240,in=120] (21) (2) to [out=240,in=120] (22) (3) to [out=240,in=120] (23) (4) to [out=240,in=120] 
(24) (5) to [out=240,in=120] (25) (6) to [out=240,in=120] (26);
%draw cluster rectabgle 
\draw [blue,dotted,thick] (-0.5,6.45) -- (8,6.45) -- (8,5.55) -- (-0.5,5.55) -- (-0.5,6.45);
\draw [blue,dotted,thick] (-0.5,5.55) -- (8,5.55) -- (8,4.65) -- (-0.5,4.65) -- (-0.5,5.55);
\draw [blue,dotted,thick] (-0.5,4.65) -- (8,4.65) -- (8,3.9) -- (-0.5,3.9) -- (-0.5,4.6);
%draw the cluster label 
\node[text width=3cm] at (\x+9,6.2) {$C_0$};
\node[text width=3cm] at (\x+9,5.2) {$C_1$};
\node[text width=3cm] at (\x+9,4.3) {$C_2$};
% draw the region lines
\draw[color=blue, dotted] (\x-0.5,3.5) -- (\x-0.5,6.6);
\draw[color=blue, dotted] (\x+0.7,3.5) -- (\x+0.7,6.6);
\draw[color=blue, dotted] (\x+2,3.5) -- (\x+2,6.6);	
\draw[color=blue, dotted] (\x+3.3,3.5) -- (\x+3.3,6.6);
\draw[color=blue, dotted] (\x+4.9,3.5) -- (\x+4.9,6.6);	
\draw[color=blue, dotted] (\x+6.3,3.5) -- (\x+6.3,6.6);	
\draw[color=blue, dotted] (\x+8,3.5) -- (\x+8,6.6);	
% dear the region labels
\node[text width=2cm] at (\x+0.9,3.6) {$R_a$};
\node[text width=2cm] at (\x+2.3,3.6) {$R_b$};
\node[text width=2cm] at (\x+3.7,3.6) {$R_c$};
\node[text width=2cm] at (\x+5.1,3.6) {$R_d$};
\node[text width=2cm] at (\x+6.5,3.6) {$R_e$};
\node[text width=2cm] at (\x+7.9,3.6) {$R_f$};
% draw the legends
\draw[thick] (-0.5, 6.7) -- (0.1, 6.7);
\node[text width=4cm] at (2.1, 6.7) {\small \textit{Intra-cluster overlay link (aCOL)}};
\draw[thick, dashed] (4, 6.7) -- (4.6, 6.7);
\node[text width=4cm] at (6.6, 6.7) {\small \textit{Intet-cluster overlay link (iCOL)}};
\end{tikzpicture}

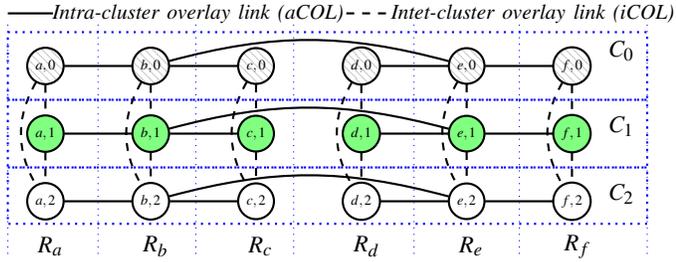
\captionof{figure}{Structured Cyclic Overlay Topology (SCOT) generated by CPUG operands shown in Fig. 2.}
\label{fig:SCOT1}
\section{Clustering for Structuredness}
Parallel paths or links provided by a SCOT are not enough to handle issues presented in Sec. 2. This section describes a set of classifications for brokers and links to build a \textit{structuredness} in SCOT. The \textit{structuredness} divides a SCOT into uniquely identifiable group of brokers called \textit{clusters}. Another pattern of grouping divides a SCOT into multiple \textit{regions}. Types for clusters, brokers, and links are used to generate subscription--trees of shortest lengths, avoid use of unique identifications to detect loops, and support inter-cluster dynamic routing (cf. Secs. 5 \& 6). More details are provided in the following.
\subsection{Cluster and Region}
Each $G^i_{af}-fiber$ in a SCOT is a separate group of brokers called a \textit{SCOT Cluster} (or simply a cluster) and represented by $C_{i}$, where $i \in V_{cf}$ is known as \textit{Cluster Index}. A cluster index is the label of a vertex of $V_{cf}$ that generates the cluster (or $G^i_{af}-fiber$) when a CPUG is calculated. Similarly, each $G^j_{cf}-fiber$ is called a \textit{Region} and is represented as $R_{j}$, where $j \in V_{af}$ is a \textit{Region Index}. A region index is the label of a vertex of $V_{cf}$ that generates the region (or $G^j_{cf}-fiber$) when a CPUG is calculated. There are $|V_{cf}|$ and $|V_{af}|$ number of clusters and regions in a SCOT, respectively. The SCOT in Fig. 3 contains three clusters (horizontal layers) each identified by $C_{i}$, where $i \in \{0, 1, 2\}$, and six regions (vertical layers) each identified by a unique $R_{j}$, where $j \in {\{a,b,c,d,e,f\}}$. 
\subsection{Overlay Links and Messaging}
A SCOT has two types of links: (i) an \textit{intra-cluster overlay link (aCOL)}, and (ii) an \textit{inter-cluster overlay link (iCOL)}. aCOLs connect brokers in the same cluster, while iCOLs connect brokers in the same region. Messaging along aCOLs and iCOLs is referred to as \textit{intra-} and \textit{inter-cluster messaging}, respectively. The set of all aCOLs in a cluster $C_{i}$ is  $\{l \langle (x, i,), (y, i) \rangle | x, y \in V_{af} \wedge xy \in E_{af}\}$, while the set of all aCOLs in a SCOT is 
\begin{center}
	$\{l \langle (x, z,), (y, z) \rangle | x, y \in V_{af} \wedge xy \in E_{af} \wedge z \in V_{cf} \}$.
\end{center}
Similarly, the set of all iCOLs in a region $R_{j}$ is $\{ l \langle (j, x'), (j, y') \rangle | x',y'  \in V_{cf} \wedge x'y' \in E_{cf} \}$, while the set of all iCOLs in a SCOT is 
\begin{center}
	$\{ l \langle (z', x'), (z', y') \rangle | x',y'  \in V_{cf} \wedge x'y' \in E_{cf} \wedge z' \in V_{af} \}$.
\end{center}
There are $|E_{af}|.|V_{cf}|$ number of aCOLs, and $|V_{af}|.|E_{cf}|$ iCOLs in a SCOT. A \textit{target link} refers to an overlay link that is part of a notification routing path. 
\subsection{Classification of Clusters and Brokers}
Classification or types of SCOT brokers and clusters is used for cluster-level routing of subscriptions and notifications (cf. Secs. 5 \& 6). The cluster that contains the host broker of a client is the \textit{Primary or Host Cluster}, while all other are the \textit{Secondary Clusters} of the client. The \textit{Primary neighbours} of a broker belong to the same cluster, while the \textit{Secondary neighbours} are those in the same region. Primary and secondary neighbours are also called direct neighbours. This arrangement of secondary brokers requires only one iCOL to forward messages from one cluster to any other cluster. The host (or secondary) cluster of a publisher is its \textit{Target Cluster} (or \textit{Target Secondary Cluster (TSC)}) if the cluster hosts at least one interested subscriber. An \textit{edge broker} has at most one primary neighbour, while an \textit{inner broker} has at least two primary neighbours. All brokers in a region are the same type (i.e., are either inner or edge). A SCOT broker is represented by $B(x, y)$, where $x \in V_{af}$ and $y \in V_{cf}$ (from Eq. 1), and is aware of its own type (i.e., edge or inner), the types of its primary and secondary neighbours, and the types of its links (i.e., aCOLs and iCOLs). 
\begin{tikzpicture}
\def\y {0.25}
% draw boxes
\foreach \x in {-0.5, 0, 0.5, 1, 1.5, 2, 2.5, 3}
\node[draw, rectangle, scale=2.1, thick] (B1)	at  (\x, 1) {};
% draw index
\node at  (-0.5, 1.5) {7};
\node at  (0, 1.5) {6};
\node at  (0.5, 1.5) {5};
\node at  (1, 1.5) {4};
\node at  (1.5, 1.5) {3};
\node at  (2, 1.5) {2};
\node at  (2.5, 1.5) {1};
\node at  (3, 1.5) {0};
% text for bits
\foreach \x in {-0.5, 0, 0.5, 1, 1.5, 2.5, 3}
\node at  (\x, 1) {0};
\node at  (2, 1) {1};
\draw [decorate,decoration={brace,amplitude=4pt}, rotate=0, thick] (-0.5, 1.7) -- (3, 1.7) node [black,midway,yshift=0.4cm] {\footnotesize \textit{Bit Indexes}};
\draw [decoration={brace,mirror,raise=5pt, amplitude=4pt},decorate, thick] (-0.7, 0.85) --  node[below=10pt]{\footnotesize \textit{Bit Values}} (3.2, 0.85) ; 
%start of the second grid....
\foreach \x in {-0.5, 0, 0.5, 1, 1.5, 2, 2.5, 3}
\node[draw, rectangle, scale=2.1, thick] (B2)	at  (\x+4.2, 1) {};
% draw index
\node at  (-0.5+4.2, 1.5) {7};
\node at  (0+4.2, 1.5) {6};
\node at  (0.5+4.2, 1.5) {5};
\node at  (1+4.2, 1.5) {4};
\node at  (1.5+4.2, 1.5) {3};
\node at  (2+4.2, 1.5) {2};
\node at  (2.5+4.2, 1.5) {1};
\node at  (3+4.2, 1.5) {0};

\foreach \x in {-0.5, 0, 0.5, 1, 1.5, 2.5}
\node at  (\x+4.2, 1) {0};

\node at  (2+4.2, 1) {1};
\node at  (3+4.2, 1) {1};

\draw [decorate,decoration={brace,amplitude=4pt}, rotate=0, thick] (-0.5 + 4.2, 1.7) -- (3 + 4.2, 1.7) node [black,midway,yshift=0.4cm] {\footnotesize \textit{Bit Indexes}};
\draw [decoration={brace,mirror,raise=5pt, amplitude=4pt},decorate, thick] (-0.7+4.2, 0.85) --  node[below=10pt]{\footnotesize \textit {Bit Values}}(3.2+4.2, 0.85);

\node at (1.2, -0.2) 		  	{\small (a) $CBV_{s}$ for a subscription.};
\node at (5.5, -0.2) 			{\small (b) $CBV_{p}$ for a notfication.};

\end{tikzpicture}

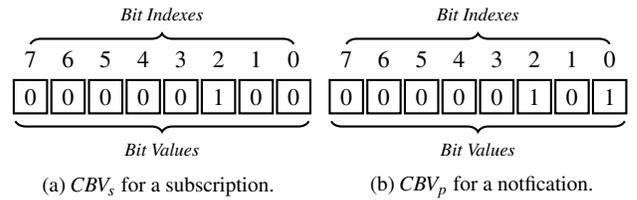
\captionof {figure}{(a) the $CBV_{s}$ of a subscription indicates that the subscriber is hosted by a broker in cluster $C_{2}$. (b) the $CBV_{p}$ of a notification indicates that $C_{0}$ and $C_{2}$ are the secondary target clusters.}
\label{fig:SRA2}
\subsection{Cluster Bit Vector}
Cluster Bit Vector (CBV) is a row (vector) of bits used to identify the host cluster of a subscriber and the TSCs of a publisher. It has two contexts: (i) \textit{the subscription context} $CBV_{s}$ identifies host cluster of the subscriber, and (ii) \textit{the publication context} $CBV_{p}$ identifies TSCs in inter-cluster dynamic routing (cf. Sec. 6). Bits in CBV are indexed from right to left, with the index of the right most bit being zero. Each bit of CBV is reserved for a SCOT cluster where index of the bit is same as index of the cluster it represents. Since each subscriber has at most one host cluster, there is only one \textit{meaningful} bit in each $CBV_{s}$. The $CBV_{s}[2]$ in Fig. 4(a) indicates that $C_{2}$ is the host cluster of the subscriber (index of the bit and cluster is 2). Each broker is aware of index bit of its cluster and set it to 1 before a local subscription is broadcast. The $CBV_{p}$ of a notification in Fig. 4(b) indicates that $C_{0}$ and $C_{2}$ are the TSCs of the publisher and should receive the notification. The number of significant bits in a CBV is equal to the number of clusters in a SCOT. The SCOT in Fig. 3 has three clusters and requires only three bits in CBV to identify all possible TSCs. \texttt{OctopiS} saves $CBV_{s}$ in routing tables while $CBV_{p}$ is carried with a notification in inter-cluster dynamic routing.
\subsection{Example}
We use Fig. 2 and Fig. 3 to explain structure of a clustered SCOT. Fig. 2 shows that $|G_{af}| = 6$ (i.e.,  $||G_{af}|| = 5$), while $|G_{cf}| = 3$ and $||G_{cf}|| = 3$. There are 3 clusters and 6 regions where each cluster is identified by $C_{i}$, where $i \in \{0,1,2\}$. Each region is identified by $R_{j}$, where $j \in \{a, b, c, d, e, f\}$. The set of all aCOLs in $C_{0}$ (or $G^0_{af}-fiber$) is given as.

$\{l \langle (a, 0,), (b, 0) \rangle | a, b \in V_{af} \wedge ab \in E_{af}\}$.\\ 
Similarly, the set of iCOLs in $R_{a}$ (or $G^a_{cf}-fiber$), is given by.

$\{l \langle (a, x',), (a, y') \rangle | x', y' \in V_{cf} \wedge x'y' \in E_{cf} \}$.\\ 
All brokers in regions $R_{a}, R_{c}, R_{d},$ and $R_{f}$ are edge, while in $R_{b}$, and $R_{e}$ are inner brokers. \textit{B(a,0)}, and \textit{B(c,0)} are the primary, while \textit{B(b,1)} and \textit{B(b, 2)} are the secondary neighbours of \textit{B(b,0)}. 
$\big(|E_{af}||V_{cf}|$ + $|V_{af}||E_{cf}|\big)$ is 33.
\section{Subscription Broadcast}
The \textit{Subscription Broadcast Process (SBP)} in traditional CPS systems is a one--step process in which a subscription reaches every broker of a cyclic overlay. However, \texttt{OctopiS} performs SBP in two--steps. We use this approach to exploit structuredness of SCOT for a \textit{controlled} SBP to generate subscription--trees of shortest--lengths, avoid using unique identification for each subscription, and prevent loops and extra IMs (eliminating \textbf{I1}, \textbf{I2}, and \textbf{I3}). Each broker of a subscriber's host cluster performs the two--steps, while the host broker of the subscriber sets its cluster index bit in $CBV_{s}$ to 1 before forwarding a subscription. $CBV_{s}[i]$ for S1, S2, and S3 in Fig. 5 are $CBV_{s}[0]$, $CBV_{s}[0]$, and $CBV_{s}[1]$, respectively. 

Each subscription in clustered SCOT has two states: (i) \textit{primary state}, and (ii) \textit{secondary state}. In the first step, a subscription is forwarded to brokers in a subscriber's host cluster and state of the subscription is primary. No loops occur, as the host cluster is a replica of an acyclic factor $G_{af}$ (recall the acyclic property of SCOT graph operand) and the subscription broadcast is similar to in an acyclic overlay, which generates subscription--tree of the minimum length. No unique identification is needed, and no duplicates appear. aCOLs of the host cluster are added in the subscription--tree, which has the maximum length $\le diam(G_{af})$. In the second step, each broker in the host cluster of the subscriber changes state of the subscription to secondary and forwards it to secondary neighbours. The secondary neighbours do not forward a secondary subscription to any other broker. In this step, all iCOLs are added to the subscription--tree and the maximum length is $\le \big(diam(G_{af}) + 1 \big)$. A subscription with its $CBV_{s}$ is saved as \textit{\{subscription, last hop, $CBV_{s}$\}} tuple in routing tables. A broker can find state of the subscription by examining $CBV_{s}$ saved with the subscription. For a primary subscription, the index of the bit with value 1 should be same as the cluster (or broker) index. As shown in Fig. 5, in the first step of SBP for S1, \textit{B(a,0)} forwards the subscription to the primary broker \textit{B(b,0)}, and in the second step, to the secondary brokers \textit{B(a,1)} and \textit{B(a, 2)}. Each broker of $C_{0}$ repeats the first and second steps to generate the subscription--tree of S1. Similarly, the subscription--trees of S2 and S3 are generated. Contrary to traditional CPS systems, SBP in \texttt{OctopiS} is \texttt{controlled}, which always generates a unique subscription--tree for a subscription issued from a broker. Each broker is aware of its type (i.e, primary or secondary), and each cluster is treated as an exclusive acyclic overlay. SBP uses this information to forward a subscription onto specific links to brokers (e.g., in Fig. 5). Uneven load in brokers or links does not effect structure or length of a subscription--tree. This pattern of subscription broadcast does not generate duplicates.

Algorithm 1 provides more details about the two-steps of SBP. The \textit{state} attribute of a subscription has two values: \textit{PRIMARY} and \textit{SECONDARY}. The host broker of a subscriber sets the host cluster index bit $CBV_{s}[i]$ to 1 (lines 4-7), and forwards the subscription to direct neighbours. The \textit{isPrimary(n)} method checks the type of the next broker and the state attribute of the subscription is set accordingly (lines 10-11). The subscription message is saved on the primary and secondary brokers (line 15).
\begin{tikzpicture}
\def\scaleSCOT {0.6}
\def\scaleBox {0.8}
%define line/arrow styles...
\tikzstyle{line} = [draw, -latex']
\tikzstyle{lineR} = [draw, latex-']
\def\y {6}
\def\x {0}
\def\xInc {1.15}
\def\yInc {1.2}		
\node[draw, circle, scale=\scaleSCOT, pattern=north west lines, pattern color=gray!40] (1) at (\x,\y) 			{$a,0$};
\node[draw, circle, scale=\scaleSCOT, pattern=north west lines, pattern color=gray!40] (2) at (\x + \xInc*1,\y) 	{$b,0$};
\node[draw, circle, scale=\scaleSCOT, pattern=north west lines, pattern color=gray!40] (3) at (\x+\xInc*2,\y)	{$c,0$};
\node[draw, circle, scale=\scaleSCOT, pattern=north west lines, pattern color=gray!40] (4) at (\x+\xInc*3,\y) 	{$d,0$};
\node[draw, circle, scale=\scaleSCOT, pattern=north west lines, pattern color=gray!40] (5) at (\x+\xInc*4,\y) 	{$e,0$};
\node[draw, circle, scale=\scaleSCOT, pattern=north west lines, pattern color=gray!40] (6) at (\x+\xInc*5,\y) 	{$f,0$};
\node[draw, circle, scale=\scaleSCOT, fill=green!50] (11) at (\x,\y-\yInc) {$a,1$};
\node[draw, circle, scale=\scaleSCOT, fill=green!50] (12) at (\x+ \xInc*1,\y-\yInc) {$b,1$};
\node[draw, circle, scale=\scaleSCOT, fill=green!50] (13) at (\x+ \xInc*2,\y-\yInc) {$c,1$};
\node[draw, circle, scale=\scaleSCOT, fill=green!50] (14) at (\x+ \xInc*3,\y-\yInc) {$d,1$};
\node[draw, circle, scale=\scaleSCOT, fill=green!50] (15) at (\x+ \xInc*4,\y-\yInc) {$e,1$};
\node[draw, circle, scale=\scaleSCOT, fill=green!50] (16) at (\x+ \xInc*5,\y-\yInc) {$f,1$};
%draw the curved line...
\node[draw, circle, scale=\scaleSCOT] (21) at (\x,\y-\yInc*2) {$a,2$};
\node[draw, circle, scale=\scaleSCOT] (22) at (\x+ \xInc*1,\y-\yInc*2) {$b,2$};
\node[draw, circle, scale=\scaleSCOT] (23) at (\x+ \xInc*2,\y-\yInc*2) {$c,2$};
\node[draw, circle, scale=\scaleSCOT] (24) at (\x+ \xInc*3,\y-\yInc*2) {$d,2$};
\node[draw, circle, scale=\scaleSCOT] (25) at (\x+ \xInc*4,\y-\yInc*2) {$e,2$};
\node[draw, circle, scale=\scaleSCOT] (26) at (\x+ \xInc*5,\y-\yInc*2) {$f,2$};
%draw both the clients S2 and S3
\node[draw, rectangle, scale=\scaleBox, text=red] (S1)	at  (\x-1,\y) {$S1$};
\node[draw, rectangle, scale=\scaleBox] (S2)	at (\x+ \xInc*6,\y) {$S2$};
\node[draw, rectangle, scale=\scaleBox, text=blue] (S3)	at (\x+ \xInc*6, \y-\yInc) {$S3$};
%\node[draw, rectangle, scale=\scaleBox] (S4) at  (\x-1,\y-\yInc*2) {$S4$};
%draw the curved line...
%\path [line, thick] (22) to [out=30,in=150] (25);
\draw [line, red, thick] (1) to [out=30,in=150] (2);
\draw [line, thick] (6) to [out=210,in=330] (5);
\draw [line, red, thick] (2) to [out=30,in=150] (3);
\draw [line, thick] (2) to [out=210,in=330] (1); 
\draw [line, red, thick] (5) to [out=30,in=150] (6);
\draw [line, red, thick] (5) to [out=210,in=330] (4); 
\draw [line, thick] (5) -- (4);
\draw [line, thick] (2) -- (3); 
\draw [line, red, thick] (2) to [out=25,in=155] (5);
\draw [line, thick] (5) to [out=210,in=335] (2);
\draw [line, thick] (S2) to [out=150,in=30] (6);
\draw [line, red, thick] (S1) to [out=30,in=150] (1);

\path [line, red, dashed, thick] (1) to [out=260,in=90] (11);
\path [line, red, dashed, thick] (1) to [out=245,in=115] (21);
\path [line, dashed, thick] (1) to [out=245,in=110] (11);
\path [line, dashed, thick] (1) to [out=295,in=65] (21);

\path [line, red, dashed, thick] (2) to [out=260,in=90] (12);
\path [line, red, dashed, thick] (2) to [out=245,in=115] (22);
\path [line, dashed, thick] (2) to [out=245,in=110] (12);
\path [line, dashed, thick] (2) to [out=295,in=65] (22);

\path [line, red, dashed, thick] (3) to [out=260,in=90] (13);
\path [line, red, dashed, thick] (3) to [out=245,in=115] (23);
\path [line, dashed, thick] (3) to [out=245,in=110] (13);
\path [line, dashed, thick] (3) to [out=295,in=65] (23);

\path [line, red, dashed, thick] (4) to [out=260,in=90] (14);
\path [line, red, dashed, thick] (4) to [out=245,in=115] (24);
\path [line, dashed, thick] (4) to [out=245,in=110] (14);
\path [line, dashed, thick] (4) to [out=295,in=65] (24);

\path [line, red, dashed, thick] (5) to [out=260,in=90] (15);
\path [line, red, dashed, thick] (5) to [out=245,in=115] (25);
\path [line, dashed, thick] (5) to [out=245,in=110] (15);
\path [line, dashed, thick] (5) to [out=295,in=65] (25);

\path [line, red, dashed, thick] (6) to [out=260,in=90] (16);
\path [line, red, dashed, thick] (6) to [out=245,in=115] (26);
\path [line, dashed, thick] (6) to [out=245,in=110] (16);
\path [line, dashed, thick] (6) to [out=295,in=65] (26);

\draw  (S1) -- (1) (S2) -- (6); 
\draw  (S3) -- (16); 
%draw spanning tree for S3
\path [line, blue, dashed, thick] (11) to [out=70,in=275] (1);
\path [line, blue, dashed, thick] (12) to [out=70,in=275] (2);
\path [line, blue, dashed, thick] (13) to [out=70,in=275] (3);
\path [line, blue, dashed, thick] (14) to [out=70,in=275] (4);
\path [line, blue, dashed, thick] (15) to [out=70,in=275] (5);
\path [line, blue, dashed, thick] (16) to [out=70,in=275] (6);
\path [line, blue, dashed, thick] (11) to [out=290,in=85] (21);
\path [line, blue, dashed, thick] (12) to [out=290,in=85] (22);
\path [line, blue, dashed, thick] (13) to [out=290,in=85] (23);
\path [line, blue, dashed, thick] (14) to [out=290,in=85] (24);
\path [line, blue, dashed, thick] (15) to [out=290,in=85] (25);
\path [line, blue, dashed, thick] (16) to [out=290,in=85] (26);
%S3, in the same cluster
\draw [line, blue, thick] (16) to [out=210,in=330] (15); 
\draw [line, blue, thick] (15) to [out=210,in=330] (14);
\draw [line, blue, thick] (12) to [out=330,in=210] (13);
\draw [line, blue, thick] (12) to [out=210,in=330] (11);
\draw [line, blue, thick] (S3) to [out=160,in=20] (16);
%draw the curved arrow...
\path [line, blue, thick] (15) to [out=155,in=30] (12);
\end{tikzpicture}
\captionof{figure}{The two--step subscription forwarding process in the SCOT in Fig. 3. $C_{0}$ is the host cluster of S1 and S2, while $C_{1}$ is the host cluster of S3. The solid arrows indicate part of the subscription--trees generated in the first step and the dashed arrows indicate part of the subscription--trees generated in the second step.\\}
\label{fig:theSFP} 
\begin {algorithm}
\KwIn{$s:$ a subscription message\;}
\KwOut{$DL:$ a list of next destinations that should receive $s$ \;}
\BlankLine
\textit{/* PRIMARY state indicates host cluster of the subscriber */}\;
\If{$ s.state = PRIMARY \vee isHostBroker (s)$}{
	\BlankLine
	\textit{/* Host broker sets the index bit for the subscription. i is the cluster index. */}\;
	\If{$isHostBroker(s)$}{
		$s.CBV_{s}[i] \gets setBitOn(i)$\;
		$s.hostBroker \gets false$\;
		$s.state \gets PRIMARY$\;	
	}
	%send this sub message to all primary neighbours
	\BlankLine
	\textit{/* forward subscription to all primary neighbours */}\;
	\ForEach {$n \in (NeighbourList - Sender)$}{
		\If{$isPrimary(n) \neq true$}{
			$s.state \gets SECONDARY$\;			
		}
		$s.next \gets n$\;
		$DL.add(s)$\;		
	}
}
\textit{/* subscription is in SECONDARY state, not forwarded to any broker */}\;
$RT.insert(s)$ \;	
\BlankLine		
\label{algo:A1}
\caption{$scotSBP(s)$}
\end{algorithm}
The subscription--trees in clustered SCOT require no updates in secondary brokers when a subscriber relocates to some other broker in the same cluster as the second step of SBP does not change. This provides robust fault tolerance when a broker fails for some reasons (fault tolerance in \texttt{OctopiS} is not within scope of this paper).
\section{Notification Routing}
This section describes our static and inter--cluster dynamic routing approaches in clustered SCOT. \texttt{OctopiS} uses cluster--based static routing to deliver notifications using subscription--trees of shortest lengths, and switches to dynamic routing when congestion is detected. We also outline state--of--the--art BID-based routing in unclustered SCOT.   
\subsection{BID--based Static Routing}
An unclustered SCOT has no types or groups of brokers and links, which is prerequisite to use BID--based routing algorithm. To identify routing paths in an unclustered SCOT, BID--based routing algorithm requires a notification to carry BIDs assigned to matching subscriptions. SBP in unclustered SCOT is uncontrolled and lengths of routing paths satisfy the relation:
\begin{center}
	$max\big(d\langle(x_{1},y_{1}), (x_{2}, y_{2})\rangle \big) \leq \big (|G_{cf}| (diam(G_{af}) + 1)-1 \big)$
\end{center}  
where $(x_{1},y_{1})$ and $(x_{2}, y_{2})$ are any brokers. Increase in payload due to adding BIDs is an important concern because normally a BID is formed by a combination of IP address and a broker level unique identifier \cite{PADRESBookChapte}. For a large network, where a large number of brokers may host interested subscribers, BID--based routing is not scalable because of additional payload and impeding in-broker processing in matching process \cite{carz_thesis, Li_ADAP}.\\ 
As Algorithm 2 shows, the host broker of a publisher adds BIDs of the matching subscriptions in \textit{bidList} attribute of a notification $n$ (lines 3-7). \textit{splitBIDs} method creates hash map, which uses a next destination path as hash key and the list of BIDs of brokers down to the next destination path as object of the hash key. The presence of an object \textit{localBID} in the hash map indicates that each local subscriber with a matching subscriptions should receive a copy of $n$ (lines 11-16). Afterwards, a copy of $n$ with the corresponding list of the remaining BIDs are forwarded onto the next destinations (lines 18-21).
\begin {algorithm}
\KwIn{$n:$ a notification message\;}
\KwRet{$DL:$ a list of next destinations that should receive $n$\;}
\BlankLine
\textit{/* host broker of publisher adds BIDs*/}\;
\If {$isHostBroker(n)$}{
	$IS \gets getInterestedSubs(n)$\;
	%embed BIDs into pub msg.
	\ForEach {$s \in IS$}{
		$n.bidList \gets s.bid$\;
	}
	$n.hostBroker \gets false$\; 
}
\textit{/* next destination based split list of BIDs */}\;
$nextBIDs \gets splitBIDs(n.bidList)$\;
\BlankLine
\textit{/* send notifications to local interested subscribers*/}\;
\If{$localBID \in nextBIDs.objects$}{
	$localSubs \gets getLocalSubs(n)$\;
	\ForEach {$ client \in localSubs$}{
		$n.next \gets client$\;
		$DL.add(n)$\;
	}
	$nextBIDs \gets (nextBIDs - localBID)$\;
}
\BlankLine
\textit{/* send notifications to next brokers in routing paths */}\;
\ForEach {$nextHop \in nextBIDs.keys$}{
	$n.next \gets nextHop$\;
	$n.bidList \gets nextBIDs.get(nextHop)$\;
	$DL.add(n)$\;		
}
\caption{$pubBID(n)$}
\end{algorithm} 
\subsection{Cluster--based Static Routing}
\textit{Static Notification Routing (SNR)} algorithm uses subscription--trees of shortest lengths for routing notifications in a clustered SCOT. The algorithm deals with two scenarios. (i) The host cluster of the publisher is the only target cluster, and therefore lengths of the routing paths satisfy the relation:
\begin{center}
	$max\big(d\langle(u_{1},v_{1}), (u_{2}, v_{2})\rangle \big) \leq diam(G_{af})$
\end{center}
where $(u_{1},v_{1})$ and $(u_{2}, v_{2})$ are any brokers in the same cluster. (ii) At least one interested subscriber is hosted by a secondary cluster and therefore lengths of the routing paths satisfy the relation:
\begin{center}
	$max\big(d\langle(x_{1},y_{1}), (x_{2}, y_{2})\rangle \big) \leq \big( diam(G_{af}) + 1 \big)$
\end{center} 
where $(x_{1},y_{1})$ and $(x_{2}, y_{2})$ are any brokers in a clustered SCOT. The host broker of a publisher forwards a notification onto target aCOLs and iCOLs (recall that a target link is part of a routing path to next destination). The notification is routed to interested subscribers in the host and TSCs of the publisher using reverse path forwarding technique. No loops appear and no path identifications are required because each cluster is an acyclic overlay (eliminating \textbf{I4}). In Fig. 6, a notification from \textit{P1} propagates onto the routing path in $C_{2}$ to reach \textit{S4}. \textit{B(f,2)} creates two additional copies of the notification to forward to \textit{B(f,1)} and \textit{B(f,0)} in the TSCs. For \textit{S1} and \textit{S2} only one copy of the notification is forwarded to \textit{B(f,0)} to avoid duplicates. Similarly, notifications from \textit{P2} and \textit{P3} are forwarded. Note that \textit{P2} has no primary target cluster, while \textit{P3} has no TSC.\\
In Algorithm 3, \textit{getDistinctSubs} uses matching subscriptions to return a list of distinct next \textit{destinations} (line 5). The method \textit{getDistinctSubs} assures that only one copy of notification \textit{n} is forwarded onto a next destination link. The \textit{else} part (lines 7-9) handles intra--cluster messaging when \textit{n} is received by a broker other than the host broker of the publisher. $nextDistinct\_aCOLs$ provides a list of the next distinct destinations to forward \textit{n} in the same cluster.
\begin{tikzpicture}
\def\scaleSCOT {0.6}
\def\scaleBox {0.7}
%define line/arrow styles...
\tikzstyle{line} = [draw, -latex']
\def\y {6}
\def\x {1}
\def\xInc {1.15} 
\def\yInc {1.2}
\node[draw, circle, scale=\scaleSCOT, pattern=north west lines, pattern color=gray!40] (1) at (\x,\y) 			{$a,0$};
\node[draw, circle, scale=\scaleSCOT, pattern=north west lines, pattern color=gray!40] (2) at (\x + \xInc*1,\y) {$b,0$};
\node[draw, circle, scale=\scaleSCOT, pattern=north west lines, pattern color=gray!40] (3) at (\x+\xInc*2,\y)	{$c,0$};
\node[draw, circle, scale=\scaleSCOT, pattern=north west lines, pattern color=gray!40] (4) at (\x+\xInc*3,\y) 	{$d,0$};
\node[draw, circle, scale=\scaleSCOT, pattern=north west lines, pattern color=gray!40] (5) at (\x+\xInc*4,\y) 	{$e,0$};
\node[draw, circle, scale=\scaleSCOT, pattern=north west lines, pattern color=gray!40] (6) at (\x+\xInc*5,\y) 	{$f,0$};
\node[draw, circle, scale=\scaleSCOT, fill=green!50] (11) at (\x,\y-\yInc) {$a,1$};
\node[draw, circle, scale=\scaleSCOT, fill=green!50] (12) at (\x+ \xInc*1,\y-\yInc) {$b,1$};
\node[draw, circle, scale=\scaleSCOT, fill=green!50] (13) at (\x+ \xInc*2,\y-\yInc) {$c,1$};
\node[draw, circle, scale=\scaleSCOT, fill=green!50] (14) at (\x+ \xInc*3,\y-\yInc) {$d,1$};
\node[draw, circle, scale=\scaleSCOT, fill=green!50] (15) at (\x+ \xInc*4,\y-\yInc) {$e,1$};
\node[draw, circle, scale=\scaleSCOT, fill=green!50] (16) at (\x+ \xInc*5,\y-\yInc) {$f,1$};
%draw the curved line...
\node[draw, circle, scale=\scaleSCOT] (21) at (\x,\y-\yInc*2) {$a,2$};
\node[draw, circle, scale=\scaleSCOT] (22) at (\x+ \xInc*1,\y-\yInc*2) {$b,2$};
\node[draw, circle, scale=\scaleSCOT] (23) at (\x+ \xInc*2,\y-\yInc*2) {$c,2$};
\node[draw, circle, scale=\scaleSCOT] (24) at (\x+ \xInc*3,\y-\yInc*2) {$d,2$};
\node[draw, circle, scale=\scaleSCOT] (25) at (\x+ \xInc*4,\y-\yInc*2) {$e,2$};
\node[draw, circle, scale=\scaleSCOT] (26) at (\x+ \xInc*5,\y-\yInc*2) {$f,2$};
%draw the dashed i-COL lines...					
\draw [gray!60, dashed, thick]  (1) -- (11) (11) -- (21) (2) -- (12) (12) -- (22) (3) -- (13) (13) -- (23) (4) -- (14) (14) -- (24) 
(5) -- (15) (15) -- (25) (6) -- (16) (16) -- (26);
%draw the grey curved line...
\draw [gray!80, thick] (2) to [out=20,in=160] (5);
\draw [gray!80, thick] (12) to [out=20,in=160] (15);
\draw [gray!80, thick] (22) to [out=20,in=160] (25);		
%draw the topology grey lines...
\draw [gray!80, thick] (22) -- (23) (22) -- (21) (25) -- (26) (25) -- (24) (1) -- (2) (2) -- (3) (4) -- (5) (5) -- (6) (11) -- (12) (13) -- (12)
(14) -- (15) (15) -- (16);
%draw grey curved lines...
\draw [dashed, gray!80, thick] (1) to [out=240,in=120] (21) (2) to [out=240,in=120] (22) (3) to [out=240,in=120] (23) (4) to [out=240,in=120] (24) (5) to [out=240,in=120] (25) (6) to [out=240,in=120] (26);
%draw both the clients
\node[draw, rectangle, scale=\scaleBox, pattern=north west lines, pattern color=gray!40] (S1)	at  (\x-1.2,\y) {$S1$};
\node[draw, rectangle, scale=\scaleBox, pattern=north west lines, pattern color=gray!40] (S2)	at (\x+\xInc*6,\y) {$S2$};
\node[draw, rectangle, scale=\scaleBox, fill=green!50] (S3)	at (\x+ \xInc*6, \y-\yInc) {$S3$};
\node[draw, rectangle, scale=\scaleBox] (S4)	at  (\x-1.2,\y-\yInc*2) {$S4$};
\node[draw, rectangle, scale=\scaleBox, fill=green!50] (P2)	at  (\x+\xInc*6,\y-1.8) {$P2$};
\node[draw, rectangle, scale=\scaleBox, text=red] (P1)	at  (\x+ \xInc*6,\y-\yInc*2) {$P1$};
\node[draw, rectangle, scale=\scaleBox, text=green!40!red, fill=green!50] (P3)	at  (\x-1.2,\y-\yInc) {$P3$};
\draw  (S1) -- (1) (S2) -- (6) (S3) -- (16) (S4) -- (21) (P2) -- (16) (P1) -- (26) (P3) -- (11);
%messages by P2...
\draw [line, thick] (P2) to [out=180,in=320] (16);
\path [line, dashed, thick] (16) to [out=60,in=300] (6); 
\path [line, dashed, thick] (16) to [out=235,in=110] (26); 
\path [line, thick] (6) to [out=210,in=330] (5); 
\path [line, thick] (5) to [out=205,in=335] (2); 
\path [line, thick] (2) to [out=210,in=330] (1); 
\path [line, thick] (1) to [out=210,in=330] (S1); 
\path [line, thick] (6) to [out=330,in=210] (S2); 
\path [line, thick] (26) to [out=210,in=330] (25); 
\path [line, thick] (25) to [out=145,in=35] (22); 
\path [line, thick] (22) to [out=210,in=330] (21); 
\path [line, thick] (21) to [out=210,in=330] (S4); 
%messages from P1...
\draw [line, red, thick] (P1) to [out=150,in=30]  (26);				
\draw [line, red, thick] (16) to [out=335,in=205]  (S3);				
\path [line, red, thick] (26) to [out=150,in=30] (25); 
\path [line, red, thick] (25) to [out=155,in=25] (22); 
\path [line, red, thick] (22) to [out=150,in=30] (21); 
\path [line, red, thick] (21) to [out=150,in=30] (S4);
\path [line, red, thick, dashed] (26) to [out=60,in=300] (16); 
\path [line, red, thick, dashed] (26) to [out=130,in=230] (6); 
\path [line, red, thick] (6) to [out=30,in=150] (S2);	
\path [line, red, thick] (6) to [out=150,in=30] (5); 
\path [line, red, thick] (5) to [out=155,in=25] (2); 
\path [line, red, thick] (2) to [out=150,in=30] (1); 
\path [line, red, thick] (1) to [out=150,in=30] (S1);
%messages from P3...
\draw [line, green!40!red, thick] (P3) to [out=30,in=150] (11);
\draw [line, green!40!red, thick] (11) to [out=30,in=150] (12);
\draw [line, green!40!red, thick] (12) to [out=25,in=155] (15);
\draw [line, green!40!red, thick] (15) to [out=30,in=150] (16);
\draw [line, green!40!red, thick] (16) to [out=30,in=150] (S3);
\end{tikzpicture}
\captionof{figure}{Notification routing in SCOT shown in Fig. 3. $C_{2}$ is the host cluster of the publisher \textit{P1}, while $C_{1}$ is the host cluster of the publishers \textit{P2} and \textit{P3}. The set of interested subscribers for P1, P2, and P3 are \textit{\{S1, S2, S3, S4\}}, \textit{\{S1, S2, S4\}} and \textit{\{S3\}}, respectively. The set of target clusters of P1, P2 and P3 are $\{C_{0}, C_{1}, C_{2}\}, \{C_{0}, C_{2}\}$ and $\{C_{1}\}$, respectively. Red, black, and brown (dashed, solid) arrows indicate intra-- and inter--cluster notifcation routing for $P1$, $P2$, and $P3$, repectively.\\}
\label{fig:SRA}

\begin {algorithm}
\KwIn{$n:$ a notification message\;}
\KwRet{$DL:$ a list of next destinations that should receive $n$\;}
\BlankLine
\textit{/* forward n onto all target links */}\;
\If {$isHostBroker(n)$}{
	\textit{/* get distinct subscriptions to identify next target links */}\;
	$destinations \gets getDistinctSubs(n)$\;
	$n.hostBroker \gets false$\; 
}\Else{
\textit{/* forward n onto distinct target aCOLs */}\;
$destinations \gets nextDistinct\_aCOLs(n)$\;
}

%send messges to all unique detinations.
\ForEach {$d \in destinations$}{
	$n.next \gets d$\;
	$DL.add(n)$
}
\caption{$scotSNR(n)$}
\label{algoSPR}
\end{algorithm} 
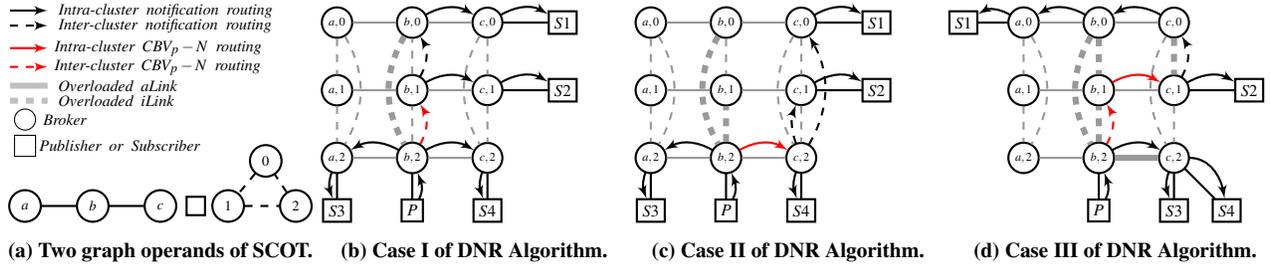
\begin{figure*}
	%define line/arrow styles...
	\tikzstyle{line} = [draw, -latex']
	\def\scaleFac {0.7}
	\centering
	\def\scaleSCOT {0.5}
	\def\scaleBox {0.7}
	\def\x {-0.5} 
	\def\xInc {1}
	\def\y {0} 
	\def\yInc {0.9}
	\def\xIncFirst {0.8}				
	\begin{subfigure}[b]{0.23\textwidth}
		\begin{tikzpicture}
		\tikzstyle{every node} = [minimum size=7mm]
		\def\y {1}
		\def\x {-1}
		\def\xInc {0.9}
		\node[draw, thick, circle, scale=0.6] (a) at (\x,\y) {$a$};
		\node[draw, thick, circle, scale=0.6] (b) at (\x+\xInc,\y) {$b$};
		\node[draw, thick, circle, scale=0.6] (c) at (\x+\xInc*2,\y) {$c$};
		%draw the curved line...
		\draw [thick] (a) -- (b) (b) -- (c);
		%draw the operator box.
		\node[draw, thick, rectangle, scale=0.35] (op)	at (\x+\xInc*3-0.43,\y) {};
		%draw the connectivity triangle...
		\node[draw, thick, circle, scale=0.6] (va) at (\x+\xInc*3,\y) 			{$1$};
		\node[draw, thick, circle, scale=0.6] (vb) at (\x+\xInc*4,\y) 			{$2$};
		\node[draw, thick, circle, scale=0.6] (vc) at (\x+\xInc*3+0.5,\y+0.6) 	{$0$};		
		%draw tirangle edges...
		\draw [dashed, thick] (va) -- (vb) (vb) -- (vc) (vc) -- (va);
		%draw the legends here ...
		\path [line, thick] (-1.2, 3.6) -- (-0.7, 3.6);
		\node[text width=3.5cm] at (1.2, 3.6) {\scriptsize \textit{Intra-cluster notification routing}};
		\path [line, thick, dashed] (-1.2, 3.4) -- (-0.7, 3.4);
		\node[text width=3.5cm] at (1.2, 3.4) {\scriptsize \textit{Inter-cluster notification routing}};
		\path [line, thick, red] (-1.2, 3.1) -- (-0.7, 3.1);
		\node[text width=4cm] at (1.4, 3.1) {\scriptsize \textit{Intra-cluster $CBV_{p}-N$ routing}};
		\path [line, thick, red, dashed] (-1.2, 2.85) -- (-0.7, 2.85);
		\node[text width=4cm] at (1.4, 2.85) {\scriptsize \textit{Inter-cluster $CBV_{p}-N$ routing}};
		\draw [line width=0.08cm, gray!50] (-1.2, 2.6) -- (-0.7, 2.6);
		\node[text width=3.5cm] at (1.2, 2.6) {\scriptsize \textit{Overloaded aLink}};
		\draw [line width=0.08cm, gray!50, dashed] (-1.2, 2.4) -- (-0.7, 2.4);
		\node[text width=3.5cm] at (1.2, 2.4) {\scriptsize \textit{Overloaded iLink}};
		\node[draw, circle, scale=0.4] (lbroker) at (-1,2.15) 			  {};
		\node[text width=1.5cm] at (0, 2.15) {\scriptsize \textit{Broker}};
		\node[draw, rectangle, scale=0.4] (lpub)	at  (-1, 1.8) 		{};
		\node[text width=3cm] at (0.7, 1.8) {\scriptsize \textit{Publisher or Subscriber}};			
		\end{tikzpicture}
		\caption{Two graph operands of SCOT.}
		\label{fig:dnr1}	
	\end{subfigure}
	~
	\begin{subfigure}[b]{0.23\textwidth}
		\begin{tikzpicture}
		%begin the first topology diagram here ...
		\tikzstyle{every node} = [thick]
		\node[draw, circle, scale=\scaleSCOT] (1) at (\x,\y+\yInc*2) {$a,0$};
		\node[draw, circle, scale=\scaleSCOT] (2) at (\x+\xInc,\y+\yInc*2) {$b,0$};
		\node[draw, circle, scale=\scaleSCOT] (3) at (\x+\xInc*2,\y+\yInc*2) {$c,0$};
		\node[draw, circle, scale=\scaleSCOT] (4) at (\x,\y+\yInc) {$a,1$};
		\node[draw, circle, scale=\scaleSCOT] (5) at (\x+\xInc,\y+\yInc) {$b,1$};
		\node[draw, circle, scale=\scaleSCOT] (6) at (\x+\xInc*2,\y+\yInc) {$c,1$};
		\node[draw, circle, scale=\scaleSCOT] (7) at (\x,\y) {$a,2$};
		\node[draw, circle, scale=\scaleSCOT] (8) at (\x+\xInc,\y) {$b,2$};
		\node[draw, circle, scale=\scaleSCOT] (9) at (\x+\xInc*2,\y) {$c,2$};
		%define the clients here...
		\node[draw, rectangle, scale=\scaleBox] (S1)	at (\x+\xInc*2+1,\y+\yInc*2)	{$S1$};
		\node[draw, rectangle, scale=\scaleBox] (P)	at (\x+\xInc,\y-0.7) 				{$P$};
		\node[draw, rectangle, scale=\scaleBox] (S2)	at (\x+\xInc*2+1,\y+\yInc)		{$S2$};
		\node[draw, rectangle, scale=\scaleBox] (S3)	at (\x,\y-0.7) 					{$S3$};
		\node[draw, rectangle, scale=\scaleBox] (S4)	at (\x+\xInc*2,\y-0.7) 			{$S4$};
		% draw striaght overlay links
		\draw [gray!80, thick] (1) -- (2) (2) -- (3) (4) -- (5) (5) -- (6) (7) -- (8) (8) -- (9) ;
		\draw [dashed, gray!80, thick] (1) -- (4) (4) -- (7) (8) -- (5) (6) -- (9) (5) -- (2) (3) -- (6);
		\draw [thick] (P) -- (8) (S1) -- (3) (S2) -- (6) (S3) -- (7) (S4) -- (9);	
		% draw curved overlay links
		\draw [dashed, thick, gray!80]  (1) to [out=300,in=60] (7);
		\draw [line width=0.08cm, dashed, gray!80]  (2) to [out=240,in=120] (8);
		\draw [dashed, gray!80, thick]  (3) to [out=240,in=120] (9);
		%arrows of P messages
		\path [line, thick] (P) to [out=60,in=290] (8);
		\path [line, dashed, thick] (5) to [out=60,in=300] (2);
		\path [line, thick] (8) to [out=30,in=150] (9);
		\path [line, thick] (5) to [out=30,in=150] (6);
		\path [line, thick] (2) to [out=30,in=150] (3);
		\path [line, thick, red, dashed] (8) to [out=60,in=300] (5);
		\path [line, thick] (6) to [out=30,in=150] (S2);
		\path [line, thick] (8) to [out=150,in=30] (7);
		\path [line, thick] (3) to [out=30,in=150] (S1);
		\path [line, thick] (9) to [out=250,in=110] (S4);
		\path [line, thick] (7) to [out=250,in=110] (S3);				
		\end{tikzpicture}
		\caption{Case I of DNR Algorithm.}
		\label{fig:dnr3}
	\end{subfigure}
	~
	\begin{subfigure}[b]{0.23\textwidth}
		\begin{tikzpicture}
		%begin the first topology diagram here ...
		\tikzstyle{every node} = [thick]
		\node[draw, circle, scale=\scaleSCOT] (1) at (\x,\y+\yInc*2) {$a,0$};
		\node[draw, circle, scale=\scaleSCOT] (2) at (\x+\xInc,\y+\yInc*2) {$b,0$};
		\node[draw, circle, scale=\scaleSCOT] (3) at (\x+\xInc*2,\y+\yInc*2) {$c,0$};
		\node[draw, circle, scale=\scaleSCOT] (4) at (\x,\y+\yInc) {$a,1$};
		\node[draw, circle, scale=\scaleSCOT] (5) at (\x+\xInc,\y+\yInc) {$b,1$};
		\node[draw, circle, scale=\scaleSCOT] (6) at (\x+\xInc*2,\y+\yInc) {$c,1$};
		\node[draw, circle, scale=\scaleSCOT] (7) at (\x,\y) {$a,2$};
		\node[draw, circle, scale=\scaleSCOT] (8) at (\x+\xInc,\y) {$b,2$};
		\node[draw, circle, scale=\scaleSCOT] (9) at (\x+\xInc*2,\y) {$c,2$};
		%define the clients here...
		\node[draw, rectangle, scale=\scaleBox] (S1)	at (\x+\xInc*2+1,\y+\yInc*2)	{$S1$};
		\node[draw, rectangle, scale=\scaleBox] (P)	at (\x+\xInc,\y-0.7) 				{$P$};
		\node[draw, rectangle, scale=\scaleBox] (S2)	at (\x+\xInc*2+1,\y+\yInc)		{$S2$};
		\node[draw, rectangle, scale=\scaleBox] (S3)	at (\x,\y-0.7) 					{$S3$};
		\node[draw, rectangle, scale=\scaleBox] (S4)	at (\x+\xInc*2,\y-0.7) 			{$S4$};
		% draw striaght overlay links
		\draw [gray!80, thick] (1) -- (2) (2) -- (3) (4) -- (5) (5) -- (6) (7) -- (8) (8) -- (9) ;
		\draw [dashed, gray!80, thick] (1) -- (4) (4) -- (7) (2) -- (5) (3) -- (6) (6) -- (9);
		\draw [thick] (P) -- (8) (S1) -- (3) (S2) -- (6) (S3) -- (7) (S4) -- (9);	
		% draw curved overlay links
		\draw [dashed, thick, gray!80]  (1) to [out=300,in=60] (7);
		\draw [line width=0.08cm, dashed, gray!80]  (2) to [out=240,in=120] (8);
		\draw [dashed, gray!80, thick]  (3) to [out=240,in=120] (9);
		\draw [line width=0.08cm, dashed, gray!80] (8) -- (5);
		%arrows of P messages
		\path [line, thick] (P) to [out=60,in=290] (8);
		\path [line, dashed, thick] (9) to [out=110,in=250] (6); 
		\path [line, dashed, thick] (9) to [out=60,in=300] (3);
		\path [line, thick, red] (8) to [out=30,in=150] (9);
		\path [line, thick] (6) to [out=30,in=150] (S2);
		\path [line, thick] (8) to [out=150,in=30] (7);
		\path [line, thick] (3) to [out=30,in=150] (S1);
		\path [line, thick] (9) to [out=250,in=110] (S4);
		\path [line, thick] (7) to [out=250,in=110] (S3);				
		\end{tikzpicture}
		\caption{Case II of DNR Algorithm.}
		\label{fig:dnr2}
	\end{subfigure}
	~
	\begin{subfigure}[b]{0.25\textwidth}
		\begin{tikzpicture}
		%begin the first topology diagram here ...
		\tikzstyle{every node} = [thick]
		\node[draw, circle, scale=\scaleSCOT] (1) at (\x,\y+\yInc*2) {$a,0$};
		\node[draw, circle, scale=\scaleSCOT] (2) at (\x+\xInc,\y+\yInc*2) {$b,0$};
		\node[draw, circle, scale=\scaleSCOT] (3) at (\x+\xInc*2,\y+\yInc*2) {$c,0$};
		\node[draw, circle, scale=\scaleSCOT] (4) at (\x,\y+\yInc) {$a,1$};
		\node[draw, circle, scale=\scaleSCOT] (5) at (\x+\xInc,\y+\yInc) {$b,1$};
		\node[draw, circle, scale=\scaleSCOT] (6) at (\x+\xInc*2,\y+\yInc) {$c,1$};
		\node[draw, circle, scale=\scaleSCOT] (7) at (\x,\y) {$a,2$};
		\node[draw, circle, scale=\scaleSCOT] (8) at (\x+\xInc,\y) {$b,2$};
		\node[draw, circle, scale=\scaleSCOT] (9) at (\x+\xInc*2,\y) {$c,2$};
		%define the clients here...
		\node[draw, rectangle, scale=\scaleBox] (S1)	at (\x-\xInc+0.2,\y+\yInc*2)	{$S1$};
		\node[draw, rectangle, scale=\scaleBox] (P)		at (\x+\xInc,\y-0.7) 			{$P$};
		\node[draw, rectangle, scale=\scaleBox] (S2)	at (\x+\xInc*2+1,\y+\yInc)		{$S2$};
		\node[draw, rectangle, scale=\scaleBox] (S3)	at (\x+\xInc*2,\y-0.7) 			{$S3$};
		\node[draw, rectangle, scale=\scaleBox] (S4)	at (\x+\xInc*2+0.7,\y-0.7) 		{$S4$};
		% draw striaght overlay links
		\draw [gray!80, thick] (1) -- (2) (2) -- (3) (4) -- (5) (5) -- (6) (7) -- (8);
		\draw [dashed, gray!80, thick] (1) -- (4) (4) -- (7) (6) -- (9);
		\draw [line width=0.08cm, gray!80] (8) -- (9);
		\draw [thick] (P) -- (8) (S1) -- (1) (S2) -- (6) (S3) -- (9) (S4) -- (9);	
		% draw curved overlay links
		\draw [dashed, thick, gray!80]  (1) to [out=300,in=60] (7);
		\draw [line width=0.08cm, dashed, gray!80]  (2) to [out=240,in=120] (8);
		\draw [dashed, gray!80, thick]  (3) to [out=240,in=120] (9);
		\draw [line width=0.08cm, dashed, gray!80] (8) -- (5) (2) -- (5) (3) -- (6);
		%arrows of P messages
		\path [line, thick] (P) to [out=60,in=290] (8);
		\path [line, dashed, thick, red] (8) to [out=60,in=300] (5); 
		\path [line, dashed, thick] (6) to [out=60,in=300] (3);
		\path [line, thick] (8) to [out=30,in=150] (9);
		\path [line, thick, red] (5) to [out=30,in=150] (6);
		\path [line, thick] (2) to [out=150,in=30] (1);	
		\path [line, thick] (3) to [out=150,in=30] (2);		
		\path [line, thick] (6) to [out=30,in=150] (S2);
		\path [line, thick] (1) to [out=150,in=30] (S1);
		\path [line, thick] (9) to [out=0,in=100] (S4);
		\path [line, thick] (9) to [out=250,in=110] (S3);				
		\end{tikzpicture}
		\caption{Case III of DNR Algorithm.}
		\label{fig:dnr4}
	\end{subfigure}
	\caption{The three cases to explain DNR algorithm. Subscribers \textit{S1}, \textit{S2}, \textit{S3}, and \textit{S4} are interested in notifications from publisher P. (b) Dynamic routing when at least one target iCOL is unoverloaded. (c) Dynamic routing when at least one target aCOL is unoverloaded. (d) Dynamic routing when all target links (aCOLs and iCOLs) are overloaded.} 
	\label{fig:DNR}
\end{figure*}
\subsection{Inter--cluster Dynamic Routing}
\textit{Dynamic routing} refers to the capability of a network system to alter a routing path in response to overloaded or failed links and/or routers. Multiple techniques have been proposed to offer dynamic routing in address--based networks where routing paths are calculated from a global view of a network topology graph that is saved on every network router \cite{routing_book}. IP addresses may be used to form clusters in a network area where the same network mask requires a single routing entry. However, these techniques are not applicable in broker--based CPS systems because brokers in these systems are aware of only their direct neighbours and destination addressing is based on contents (i.e, subscriptions). Therefore dynamic routing decisions are decentralized and have to be made without having a global view of an overlay. This section describes inter--cluster dynamic routing when one or more target inter--cluster overlay links (i.e., iCOLs) are overloaded and notifications start queuing up in the output queues. Overloading and subsequent queuing can happen in two cases: (i) when a broker is not able to process high volume of outgoing notifications and become overwhelmed, and (ii) when bandwidth is limited. As SBP generates only one subscription--tree per subscription, dynamic routing is difficult in broker-based CPS systems and never supported before. We introduce a unique approach, which deviates from traditional reverse path forwarding technique and uses structuredness of the proposed topology (i.e., SCOT) along with subscription--trees of the matching subscriptions to offer inter-cluster dynamic routing. Our approach is scalable because it does not require updates in routing tables and can reduce delivery delays when a large number of notifications start accumulating in the output queues. 

SNR algorithm adds exclusive copies of a notification in the output queues of the target links. If a publisher generates $\gamma$ number of notifications in $t_{w}$ time window, and there are $\alpha$ number of target aCOLs and $\beta$ target iCOLs, the host broker of the publisher enqueue $(\alpha + \beta).\gamma$ number of copies of notifications. A High Rate Publisher (HRP) with a high value of $\gamma$ can overwhelm brokers in a routing paths when SNR algorithm is used. Our inter-cluster \textit{Dynamic Notification Routing (DNR)} algorithm alleviates overwhelmed brokers and selects an unoverloaded iCOL to forward a notification to a TSC. The algorithm dynamically selects an unoverloaded iCOL, which may not be part of a subscription--tree to the TSC. As parallel iCOLs are available in a clustered SCOT, one can be selected without making updates in routing tables (partially eliminating \textbf{I5}). DNR adds at most \textit{one copy} of a notification in the congested output queues of a target link at a broker. The algorithm sets the cluster index bits in $CBV_{p}$ to 1 to identify those TSCs that are not forwarded the notification due to overloaded target iCOLs. $CBV_{p}$ is added in the header of the notification, called \textit{the $CBV_{p}-Notification$ (or $CBV_{p}-N$)}. $CBV_{p}-N$ is an indication for a broker that the routing is dynamic, and a copy of the notification should be forwarded to TSCs if unoverloaded target iCOLs are available. Using this technique, DNR algorithm keeps the length of the congested output queues of an overwhelmed broker to a minimum, and the load of forwarding the notification is shifted to other brokers using the heuristic that unoverloaded target iCOLs are available down the routing path. \texttt{OctopiS} uses Eq. 3 to find whether an output queue is congested, and DNR algorithm should be activated. 
\begin{equation} (Q_{\ell}) . (1+Q_{in}, 1+Q_{out})_{t_{w}}  > \tau \end{equation}
$Q_{in}$ and $Q_{out}$ are the number of notifications that enter into or leave the output queue in the time window $t_{w}$, respectively. The term $(1+Q_{in}, 1+Q_{out})_{t_{w}}$ is the ratio of $(1+Q_{in})$ to $(1+Q_{out})$, and is known as the \textit{Congestion Element (CE)}. $CE > 1$ indicates that the congestion is increased, while $CE < 1$ shows that congestion is decreased in the last $t_{w}$ interval. An output queue is congestion--free when $CE$ is 1 and the queue length $Q_{\ell}$ is 0. \texttt{OctopiS} saves the values of $Q_{in}$ and $Q_{out}$ in a \textit{Link Status Table (LST)} on each broker, and the values are updated after each $t_{w}$ interval. Inter-cluster dynamic routing by DNR algorithm is further explained in the following with help of three cases.\\
\textbf{Case I -- Unoverloaded Target iCOL}: When the output queue of at least one target iCOL is uncongested (i.e., the corresponding link is unoverloaded), DNR algorithm does not enqueue a notification in the congested output queues of the target iCOLs. Instead, only \textit{one copy} of the notification is enqueued in the uncongested output queue of the unoverloaded target iCOL. If more than one unoverloaded target iCOL is available at a broker, $CBV_{p}-N$ is forwarded onto the target iCOL, which has the least value of $Q_{\ell}$. The number of notifications added to the output queues of the target links in $t_{w}$ interval is $(\alpha + \theta ).\gamma$, where $\theta$ is the number of unoverloaded target iCOLs and $\theta < \beta$. Fig. 7(b) indicates that the iCOL $l \langle(b,2),(b,0)\rangle$), which forwards notifications from \textit{B(b,2)} to $C_{0}$, is overloaded. After finding the overloaded iCOL from the LST, \textit{B(b, 2)} sets the index bit of $C_{0}$ to 1 and adds $CBV_{p}-N$ (with \textit{$CBV_{p}$ 001}) in the output queue of the unoverloaded target iCOL $l \langle(b,2),(b,1)\rangle$) to forward $CBV_{p}-N$ to \textit{B(b, 1)}. Since the output queue of the iCOL $l \langle(b,1),(b,0)\rangle$ is uncongested, \textit{B(b, 1)} sets the index bit of $C_{0}$ to 0, removes the $CBV_{p}$ from the notification as all index bits in $CBV_{p}$ are zero, and forwards the notification to \textit{B(b, 0)} and \textit{B(c, 1)}. The notification is forwarded to S2, S3, and S4 using their subscription--trees. However, to avoid overloaded link $l \langle(b,2),(b,0)\rangle$, DNR algorithm deviates from the subscription--tree of S1 and dynamically selects the iCOL $l \langle(b,1),(b,0)\rangle$ for routing without making updates in routing tables. To forward one notification from P to interested subscribers, each of SNR and DNR algorithms generate 6 IMs, although DNR algorithm excluded one overloaded link, the dynamic routing path for S1 contains 4 brokers, which is one more than the number of brokers in path adopted by SNR algorithm.\\
\textbf{Case II -- All Target iCOLs Overloaded}: When the output queues of all target iCOLs are congested and at least one output queue of a target aCOL is uncongested, DNR algorithm uses the target aCOLs to find unoverloaded target iCOLs. $CBV_{p}-N$ is added in the uncongested output queue of the target aCOL. The load of forwarding the notification to TSCs is shifted to the next primary broker. The number of notifications added to the output queues of the target links in $t_{w}$ interval is $\alpha . \gamma$ (here $\beta$ is zero as \textit{no copy} of the notification is added into the congested output queues of the target iCOLs). Fig. 7(c) indicates that the two target iCOLs, $l \langle(b,2), (b,1) \rangle$ and $l \langle(b,2), (b,0) \rangle$, are overloaded, and notifications are sent only to \textit{B(a, 2)} and \textit{B(a, 0)} in the host cluster of P. Since two unoverloaded target aCOLs are available, the target aCOL with the least $Q_{\ell}$ is selected to forward $CBV_{p}-N$ with $CBV_{p}$ 011 (presumably, aCOL $l \langle(b,2), (c,2) \rangle$ has least value of $Q_{\ell}$). As the target iCOLs are not overloaded at \textit{B(c,2)}, the notification is routed to the TSCs after removing $CBV_{p}$. To forward one notification in this case, DNR algorithm generates 4 IMs, while SNR algorithm generates 6 IMs. Furthermore, two overloaded iCOLs are excluded from the dynamically generated routing paths.
\begin {algorithm}
\KwIn{$n:$ a notification message\;}
\KwRet{$DL:$ a list of next destinations that should receive $n$\;}
\BlankLine
\textit{/* list of interested subscribers */}\;
$IS \gets getInterestedSubs(n)$
\BlankLine
\textit{/* local hosted interested subscribers */}\;
$local \gets getHostedSubscribers(IS)$ 
\BlankLine
\textit{/* send n to local hosted subscribers */}\;
\ForEach {$d \in local$}{
	$n.next \gets d$\;
	$DL.add(n)$\;
}
\BlankLine
\textit{/* next unique destinations */}\;
$\mu \gets nextUniqueDestinations(IS-local)$\;

\BlankLine
\textit{/* get overloaded iCOLs from $\mu$ unique destinations */}\;
$\eta1 \gets getOverloaded\_iCOLs(\mu)$\;
$CBV_{p} \gets 0$\;
\BlankLine
\textit{/* Set index bits of the overloaded iCOLs */}\;
\ForEach {$d \in \eta1$}{
	\If{$d.index \neq index$}{
		$CBV_{p} \gets setBitOn(CBV_{p}, d.index)$\;
	}
}
\BlankLine
$\eta2 \gets getLeastLoaded\_iCOL(\mu)$\;
$\ell \gets getLeastLoaded\_aCOL(\mu)$\;
\BlankLine

\textit{/* Send n to unoverloaded target links (aCOLs and iCOLs) */}\;
\ForEach {$d \in (\mu - \eta1) $}{
	$n.next \gets d$\;
	$DL.add(n)$\;
}{
\BlankLine
\textit{/* Set the $CBV_{p}-N$: Cases: (i), (ii), and (iii) */}

\If{$isOverloaded(\eta2) = true \wedge isOverloaded(\ell) = false$}{
	\ForEach{$msg\in DL$}{
		\If{$msg.next = \ell$}{
			$msg.CBV_{p} \gets CBV_{p}$\;
		}
	}
}
\ElseIf{$isOverloaded(\eta2) = false$}{
	\ForEach{$msg\in DL$}{
		\If{$msg.next = \eta2$}{
			$msg.CBV_{p} \gets CBV_{p}$\;
		}
	}
}\Else{
$n.next \gets \eta2$\;
$CBV_{p} \gets setBitOFF(CBV_{p}, \eta2.index)$\;
$n.CBV_{p} \gets CBV_{p}$\;
$DL.add(n)$\;
}
}
\caption{$scotDNR(n)$}
\end{algorithm}
\\
\textbf{Case III -- All Target Links Overloaded}: If all target links are overloaded, $CBV_{p}-N$ is forwarded onto the least loaded target iCOL. Because of the possibility of having more overloaded aCOLs down the routing path, $CBV_{p}-N$ is not added into the output queue of a target aCOL even if it is the least overloaded link. The number of notifications added to the output queues of the target links in $t_{w}$ interval is $(\alpha + 1).\gamma$. Fig. 7(d) shows that $CBV_{p}-N$ is forwarded onto (presumably) the least overloaded link $l \langle(b,2),(b,1)\rangle$ with $CBV_{p}$ is 001. As the target iCOL $l \langle(b,1),(b,0)\rangle$ is also overloaded, \textit{B(b, 1)} forwards $CBV_{p}-N$ onto the target aCOL $l \langle(b,1),(c,0)\rangle$. The overloaded link $l \langle(c,1),(c,0)\rangle$ is the only target iCOL available to forward the notification to TSC $C_{0}$, DNR algorithm is unable to find unoverloaded target iCOL and the notification has to be forwarded onto the overloaded iCOL $l \langle(c,1),(c,0)\rangle$. The number of IMs generated by SNR and DNR algorithms in this case are 6 and 5 respectively. Although DNR algorithm successfully avoided the overloaded iCOL $l \langle(b,2),(b,0)\rangle$, the notification has to be forwarded onto another overloaded target iCOL $l \langle(c,1),(c,0)\rangle$. Additionally, the generated dynamic routing path for S1 has 3 additional brokers.\\
Algorithm 4 provides further details about the inter-cluster dynamic notification routing in \texttt{OctopiS}. Upon receiving a notification $n$, the overwhelmed broker prepares a list of local subscribers interested in $n$ (line 5). A copy of $n$ for each subscriber is added in the next destinations list (lines 7-9). Next, $\eta1$, a list of overloaded target iCOLs is prepared to set index bits in $CBV_{p}$ (lines 13-18). Note that the index bit of the current notification routing cluster is not set (line 17) because this information does not effect inter-cluster dynamic routing. A copy of $n$ for each unoverloaded target link is added in the next destination list \textit{DL} (lines 22-24). In the end, the algorithm shows how three cases are handled. The first \textit{if--statement} handles the Case II when all target iCOLs are overloaded and an unoverloaded target aCOL is available (lines 26-29). The second condition (in the \textit{else--if} block) is valid when an unoverloaded target iCOL is available to carry $CBV_{p}$ (lines 30-33). The \textit{else} block handles the case when all target links are overloaded and a copy of \textit{n} is added in \textit{DL} to sent onto least overloaded target iCOL $\eta2$.

DNR is a best--effort algorithm and depends on the subscription--trees laid--on by interested subscribers in a target cluster. The algorithm does not guarantee finding an unoverloaded target iCOL, even if one exists. For example, in Fig. 7(d), DNR algorithm does not use the unoverloaded link $l\langle(b,2),(a,2)\rangle$ for inter-cluster dynamic routing because the link is not a target aCOL. Forwarding the $CBV_{p}-N$ on such links can generate loops among different clusters \cite{OctopiA_TR}. DNR algorithm currently does not support \textit{intra--cluster dynamic routing}, and this is the reason that \texttt{OctopiS} eliminates \textbf{I5} only partially. \textit{intra--cluster dynamic routing} is part of the future work.
\section{Evaluation}
We implemented SNR, and DNR algorithms in \texttt{OctopiS}, developed on top of a publish/subscribe tool PADRES \cite{PADRESBookChapte}. For a comparison with state--of--the--art, we also implemented BID--based routing in PADRES. Importantly, we created a subscription--based publish/subscribe version of PADRES, as the tool supports advertisement--based semantics \cite{Li_ADAP}. SNR and DNR algorithms exploit structuredness of clustered SCOT, while BID--based routing algorithm uses unclustered SCOT.
\subsection{Setup}
Fig. 8 shows factors of the SCOT topology $\mathbb{S}_{e}$, which we used for evaluation and comparison of SNR, DNR, and BID--based routing algorithms. $G_{af}$ factor of $\mathbb{S}_{e}$ is an acyclic topology of 15 brokers (5 inner brokers (\textit{vi, vii, viii, ix} and \textit{x}) and 10 edge brokers), while $G_{cf}$ factor has 5 brokers, which generates $|V_{cf}|-1$ number of secondary neighbours for each broker in $G_{af}$. This results in 25 inner brokers and 50 edge brokers (for a total of 75 brokers), forming 5 clusters and 15 regions in $\mathbb{S}_{e}$. \texttt{OctopiS} was deployed on a cluster of 35 physical computing nodes, where each node had one quad core processor of 2.4 GHz with 4 GB RAM, and running 64-bit JDK on Linux OS. Each broker was loaded in a separate instance of JVM with 1 GB initial memory. One dedicated high throughput Gigabit Ethernet switch was used for connectivity. $\mathbb{S}_{e}$ was deployed in such a way that the primary and secondary neighbours of each broker were always on different computing nodes. 
\begin{center}
	\begin{tikzpicture}[thick]
	\scriptsize
	\centering
	\def\y {6}
	\def\x {0}
	\def\xInc {1}
	\def\yInc {0.9}
	\def\S {0.4}
	
	\node[draw, circle, minimum size=0.5cm] (1) at (\x + \xInc*0,\y) {$i$};
	\node[draw, circle, minimum size=0.5cm] (2) at (\x + \xInc*1,\y) {$ii$};
	\node[draw, circle, minimum size=0.5cm] (3) at (\x + \xInc*2,\y) {$iii$};
	\node[draw, circle, minimum size=0.5cm] (4) at (\x + \xInc*3,\y) {$iv$};
	\node[draw, circle, minimum size=0.5cm] (5) at (\x + \xInc*4,\y) {$v$};
	\node[draw, circle, minimum size=0.5cm] (11) at (\x + \xInc*0,\y-\yInc) {$vi$};
	\node[draw, circle, minimum size=0.5cm] (12) at (\x + \xInc*1,\y-\yInc) {$vii$};
	\node[draw, circle, minimum size=0.5cm] (13) at (\x + \xInc*2,\y-\yInc) {$viii$};
	\node[draw, circle, minimum size=0.5cm] (14) at (\x + \xInc*3,\y-\yInc) {$ix$};
	\node[draw, circle, minimum size=0.5cm] (15) at (\x + \xInc*4,\y-\yInc) {$x$};
	\node[draw, circle, minimum size=0.5cm] (21) at (\x + \xInc*0,\y-\yInc*2) {$xi$};
	\node[draw, circle, minimum size=0.5cm] (22) at (\x + \xInc*1,\y-\yInc*2) {$xii$};
	\node[draw, circle, minimum size=0.5cm] (23) at (\x + \xInc*2,\y-\yInc*2) {$xiii$};
	\node[draw, circle, minimum size=0.5cm] (24) at (\x + \xInc*3,\y-\yInc*2) {$xiv$};
	\node[draw, circle, minimum size=0.5cm] (25) at (\x + \xInc*4,\y-\yInc*2) {$xv$};
	\draw [thick] (1) -- (11) (11) -- (12) (12) -- (13) (14) -- (15) (12) -- (22) (11) -- (21) (13) -- (23) (13) -- (14) (14) -- (24)  (15) -- (25) (2) -- (12) (3) -- (13) (4) -- (14) (5) -- (15) ;
	%draw the operator box.
	\node[draw, rectangle, minimum size=0.3cm] (op)    at (4.65,\y-\yInc) {};
	%draw the connectivity triangle...
	\node[draw, circle, minimum size=0.5cm] (c0) at (6.4,\y-0.1)	{$0$};
	\node[draw, circle, minimum size=0.5cm] (c1) at (5.3, \y-\yInc) {$1$};
	\node[draw, circle, minimum size=0.5cm] (c2) at (7.4,\y-\yInc) {$4$};
	\node[draw, circle, minimum size=0.5cm] (c3) at (5.6,\y-\yInc*2) {$2$};
	\node[draw, circle, minimum size=0.5cm] (c4) at (7,\y-\yInc*2) {$3$};
	%draw edges...
	\draw [dashed] (c0) -- (c1) (c0) -- (c2) (c0) -- (c3) (c0) -- (c4) (c1) -- (c3) (c1) -- (c2) (c1) -- (c4) (c3) -- (c4) (c3) -- (c2) (c4) -- (c2);
	\end{tikzpicture}
	
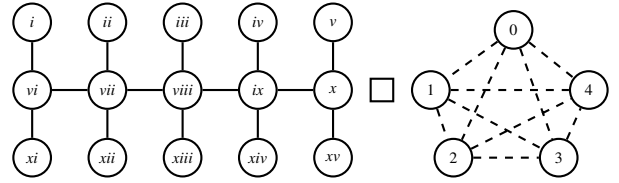
\captionof{figure}{\textit{Left operand of $\Box$ operator is the $G_{af}$ factor while the right operand is the $G_{cf}$ factor of $\mathbb{S}_{e}$. The $G_{cf}$ generates five replicas of $G_{af}$.\\}}
	\label{fig:evalTop}
\end{center}
Stock datasets are commonly used to generate workloads for evaluations of CPS systems \cite{agg15}. We used a dataset of 500 stock symbols from Yahoo Finance!, where each stock notification had 10 distinct attributes. This high dimension data require high computation for filtering information during in--broker processing. Subscriptions were generated synthetically. We randomly distributed publishers and subscribers, where each subscriber registered one subscription with 2\% selectivity.
\subsection{Metrics}
Through experiments with real world data, we evaluated \texttt{OctopiS} using primitive metrics, such as subscription, notification, and matching delays.\\
\textit{\textbf{Subscription delay}}: The subscription (forwarding) delay is the maximum time elapsed as a subscription reaches brokers in an overlay network. A subscription is expected to take less time to reach brokers in a close proximity to the host broker of a subscriber than to brokers in a far--off region. Since the SBP in a clustered SCOT generates subscription--trees of shortest lengths, it is important to measure the difference between the average subscription delays in unclustered and clustered SCOT.\\
\textit{\textbf{Notification delay}}: The notification delay measures end--to--end latency from the time a notification is generated to the time it is received by a subscriber. As the average length of subscription--trees for BID--based routing is higher than a clustered SCOT for SNR and DNR algorithms, knowing the difference in latencies in these two cases is a worthwhile area of inquiry.\\
\textit{\textbf{Matching delay}}: The matching delay is the time taken to find subscriptions that match with the notification contents. BID--based routing is expected to have less matching delay as the matching is done at only the brokers which host publishers and interested subscribers.\\
\textit{\textbf{Inter--broker Messages (IMs)}}: The number of IMs depends on the lengths of subscription--trees, as well as the relative distance between publishers and subscribers. As the average length of subscription--trees in clustered SCOT is less than in unclustered SCOT, the number of IMs generated by SNR and DNR algorithms is expected to be less than BID--based routing algorithm.

Aggregation techniques like \textit{covering} are developed for acyclic overlays to reduce size of routing tables. These techniques can be used with clustered SCOT as each cluster is an acyclic overlay. However, for a comparison with state--of--the--art BID--based routing, covering in not considered in evaluation because no covering technique for cyclic overlays is reported in literature.  
% trim={<left> <lower> <right> <upper>}
\begin{figure*}
	\centering
	\begin{subfigure}[b]{0.24\textwidth}
		\includegraphics[trim=1cm 0.4cm 0.1cm 1cm, width=0.95\textwidth]{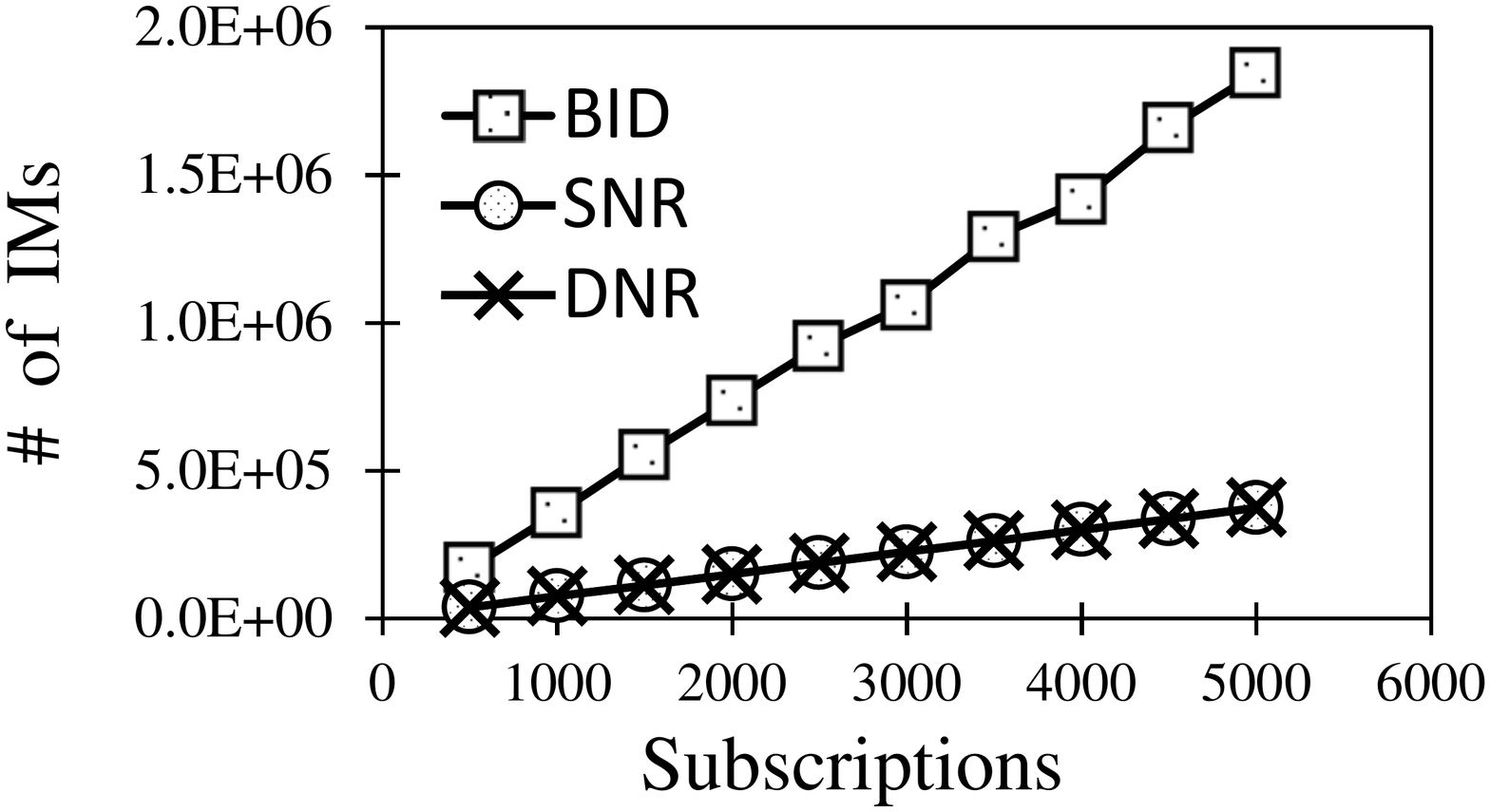}
		\caption{\small \# of IMs in SBP}
		\label{fig:S2}
	\end{subfigure}
	~
	\begin{subfigure}[b]{0.24\textwidth}
		\includegraphics[trim=1cm 0.4cm 0.1cm 1cm, width=0.9\textwidth]{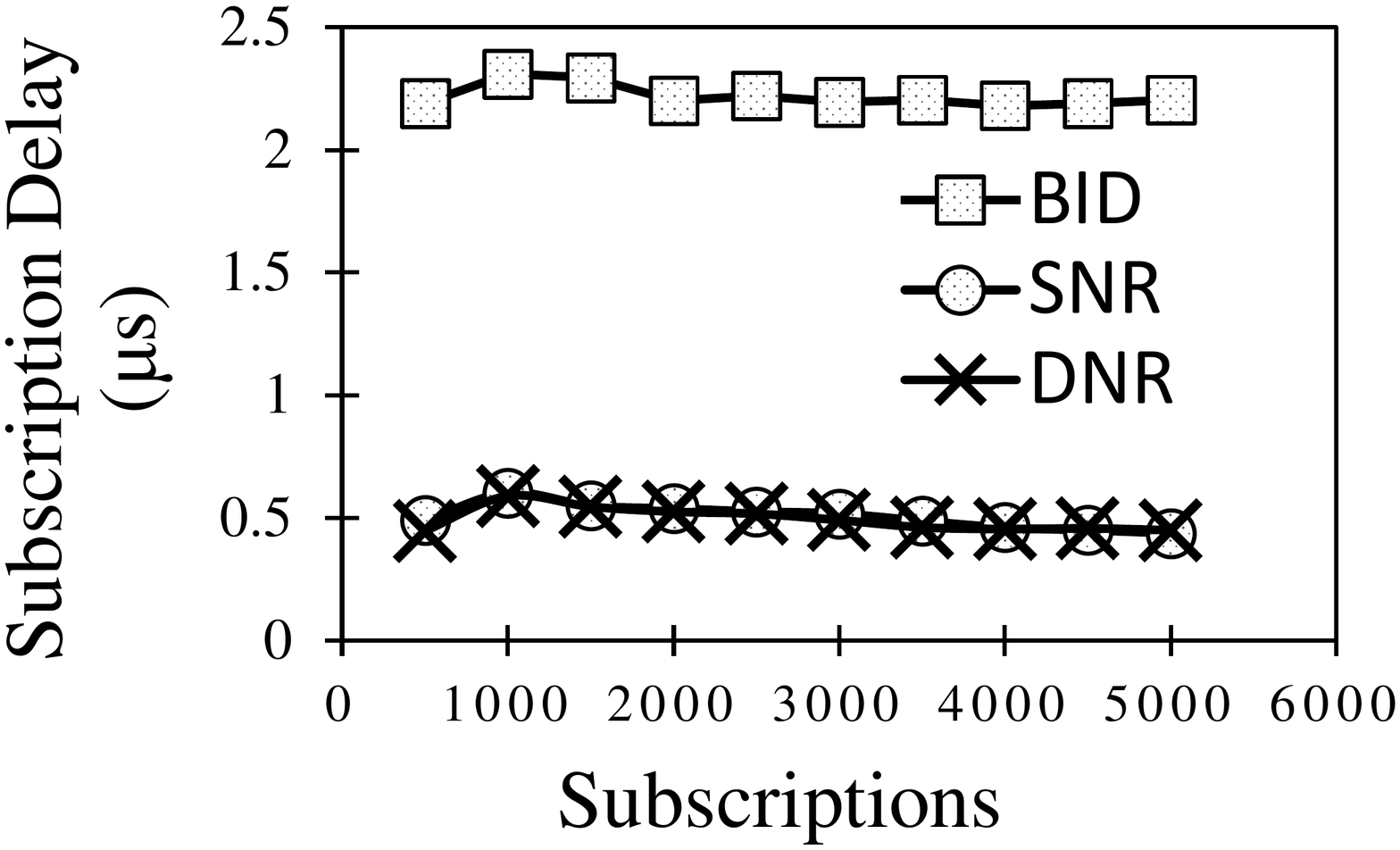}
		\caption{\small Subscription delay}
		\label{fig:S1}
	\end{subfigure}
	~ % start the next figure...
	\begin{subfigure}[b]{0.24\textwidth}
		\includegraphics[trim=1cm 0.4cm 0.1cm 1cm, width=0.88\textwidth]{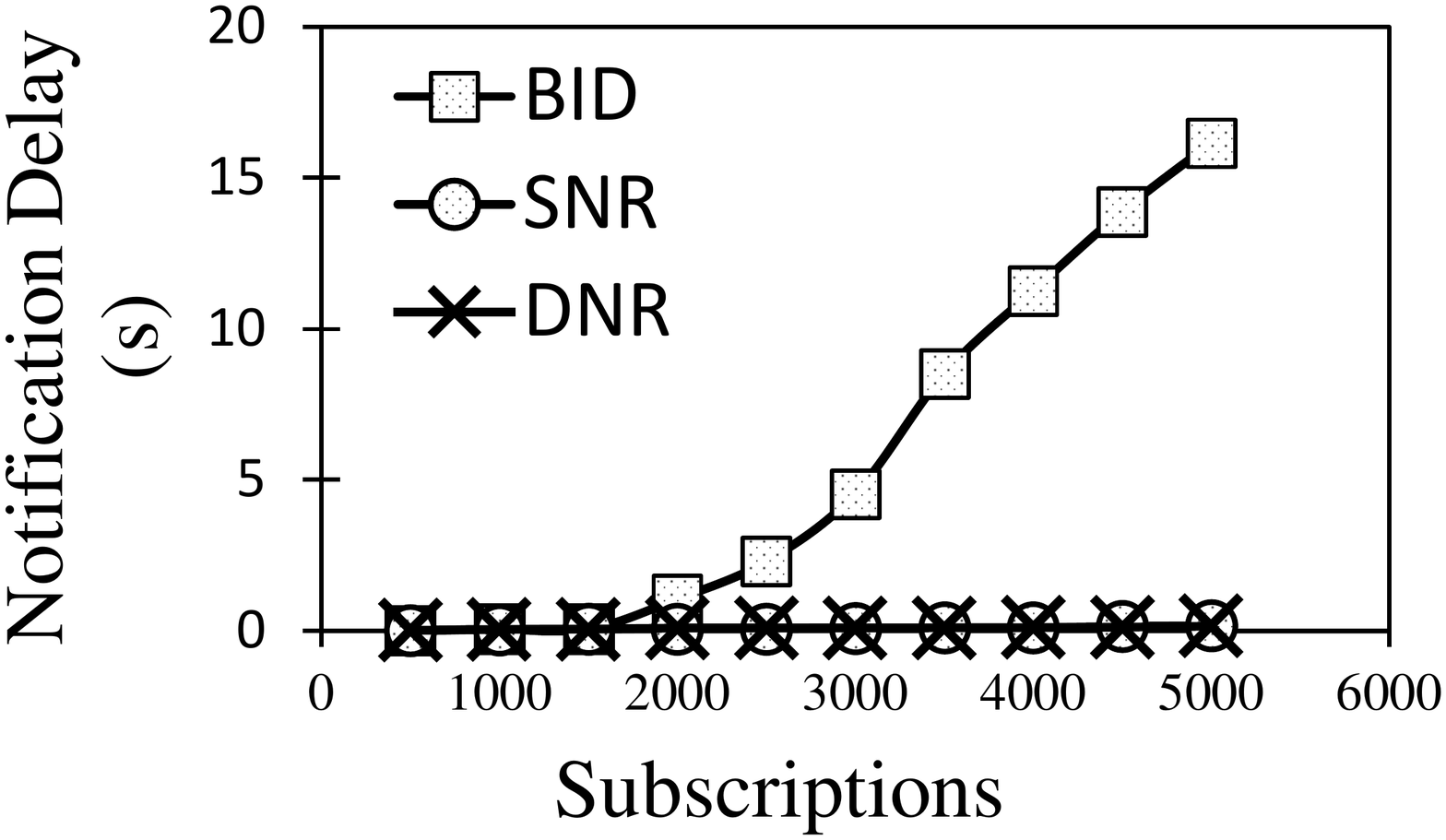}
		\caption{\small Notification delay}
		\label{fig:P1}
	\end{subfigure}
	~ %add desired spacing between images, e. g. ~, \quad, \qquad, \hfill etc.
	\begin{subfigure}[b]{0.24\textwidth}
		\includegraphics[trim=1.5cm 0.4cm 0.1cm 1cm, width=0.88\textwidth]{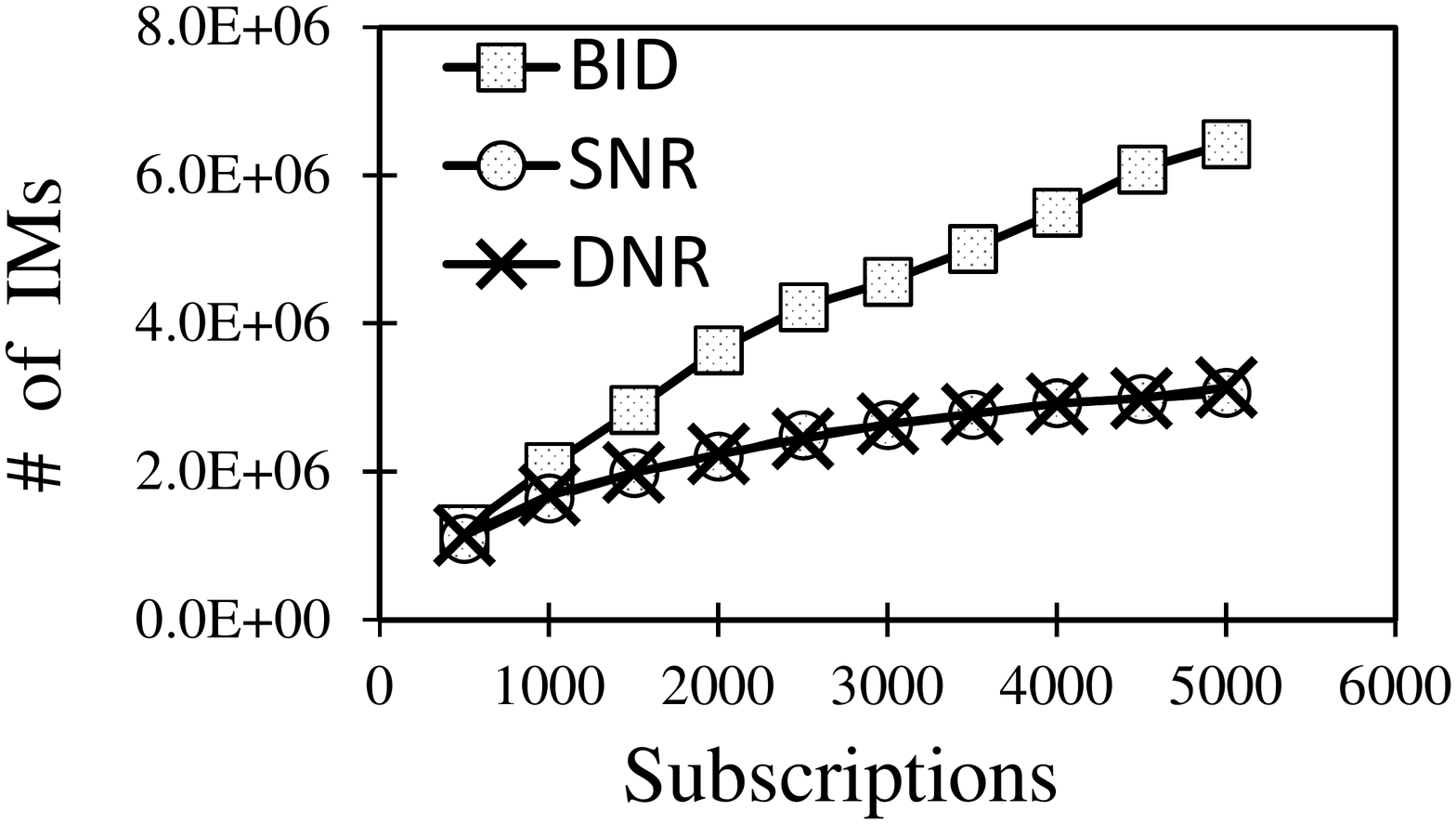}
		\caption{\small  \# of IMs in notification routing}
	\end{subfigure}
	\caption{\textit{Subscriber Scalability in an unclustered and clustered SCOT.}}\label{fig:publication}
\end{figure*}
% trim={<left> <lower> <right> <upper>}
\begin{figure*}
	\centering
	\begin{subfigure}[b]{0.24\textwidth}
		\includegraphics[trim=1cm 0.5cm 0.5cm 1cm, width=0.9\textwidth]{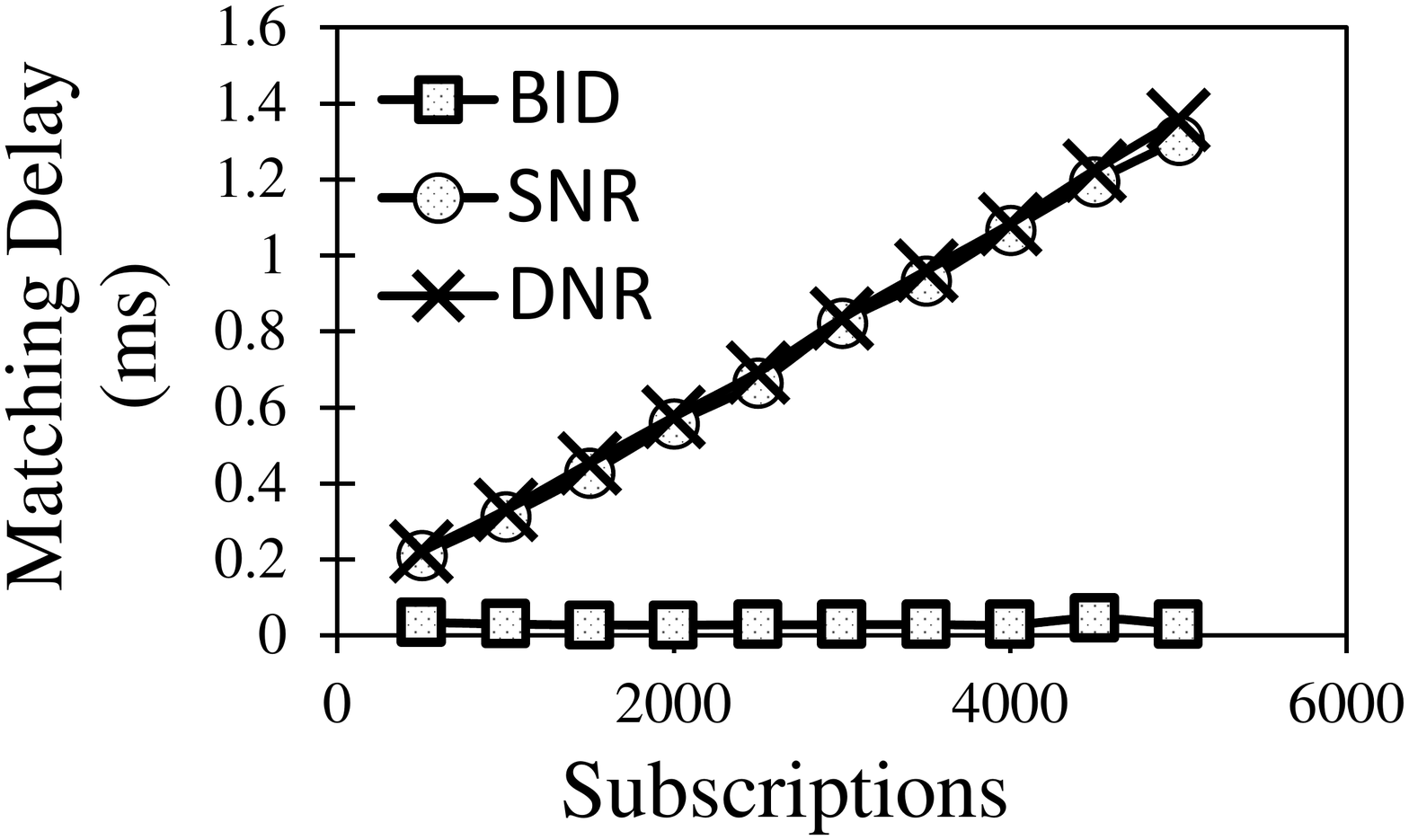}
		\caption{\small Matching delay}
		\label{fig:S55}
	\end{subfigure}
	~
	\begin{subfigure}[b]{0.24\textwidth}
		\includegraphics[trim=1cm 0.5cm 0.5cm 1cm, width=0.88\textwidth]{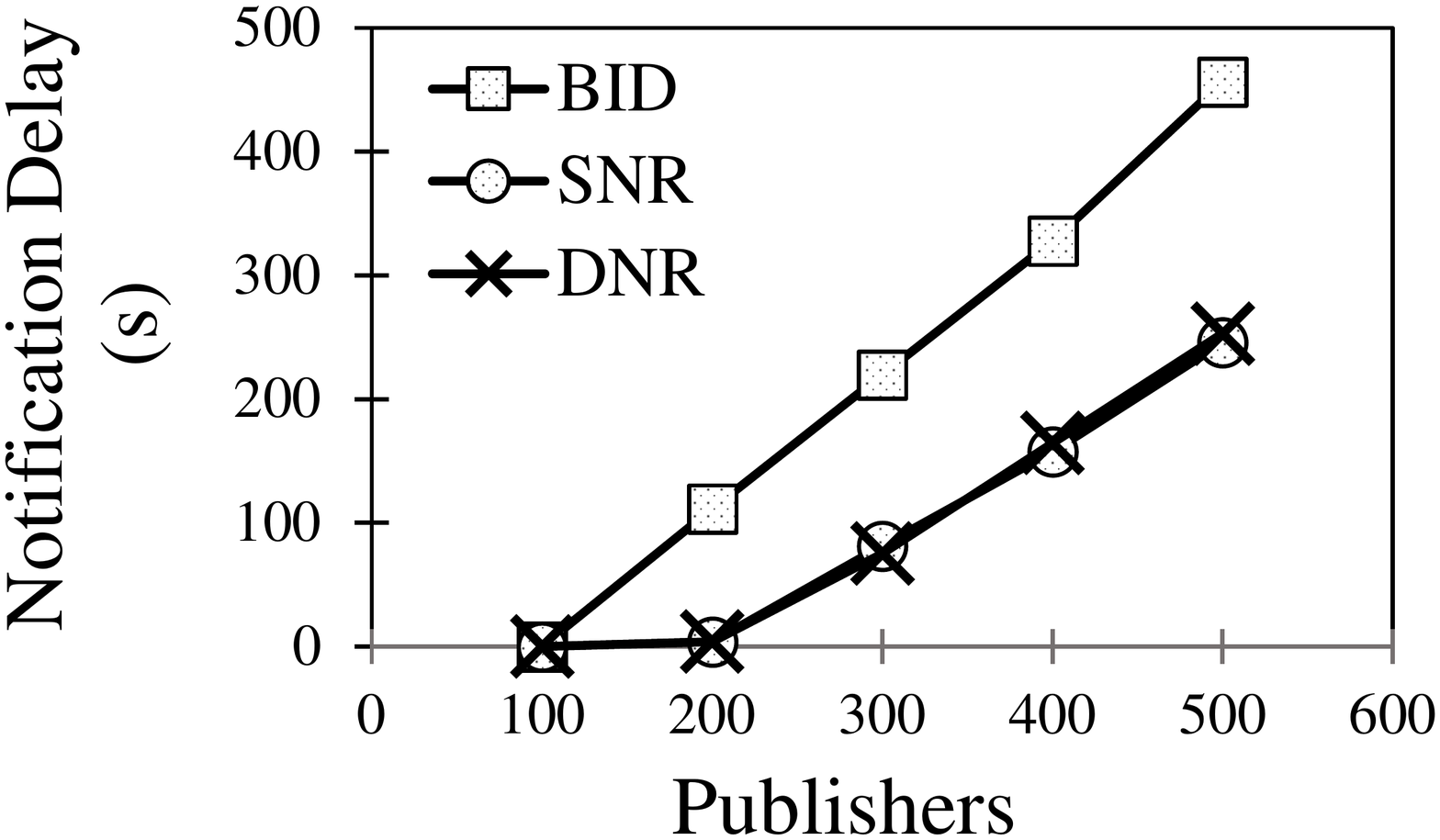}
		\caption{\small Notification delay}
		\label{fig:P4}
	\end{subfigure}
	~ %add desired spacing between images, e. g. ~, \quad, \qquad, \hfill etc.
	\begin{subfigure}[b]{0.24\textwidth}
		\includegraphics[trim=1cm 0.5cm 0.5cm 1cm, width=0.88\textwidth]{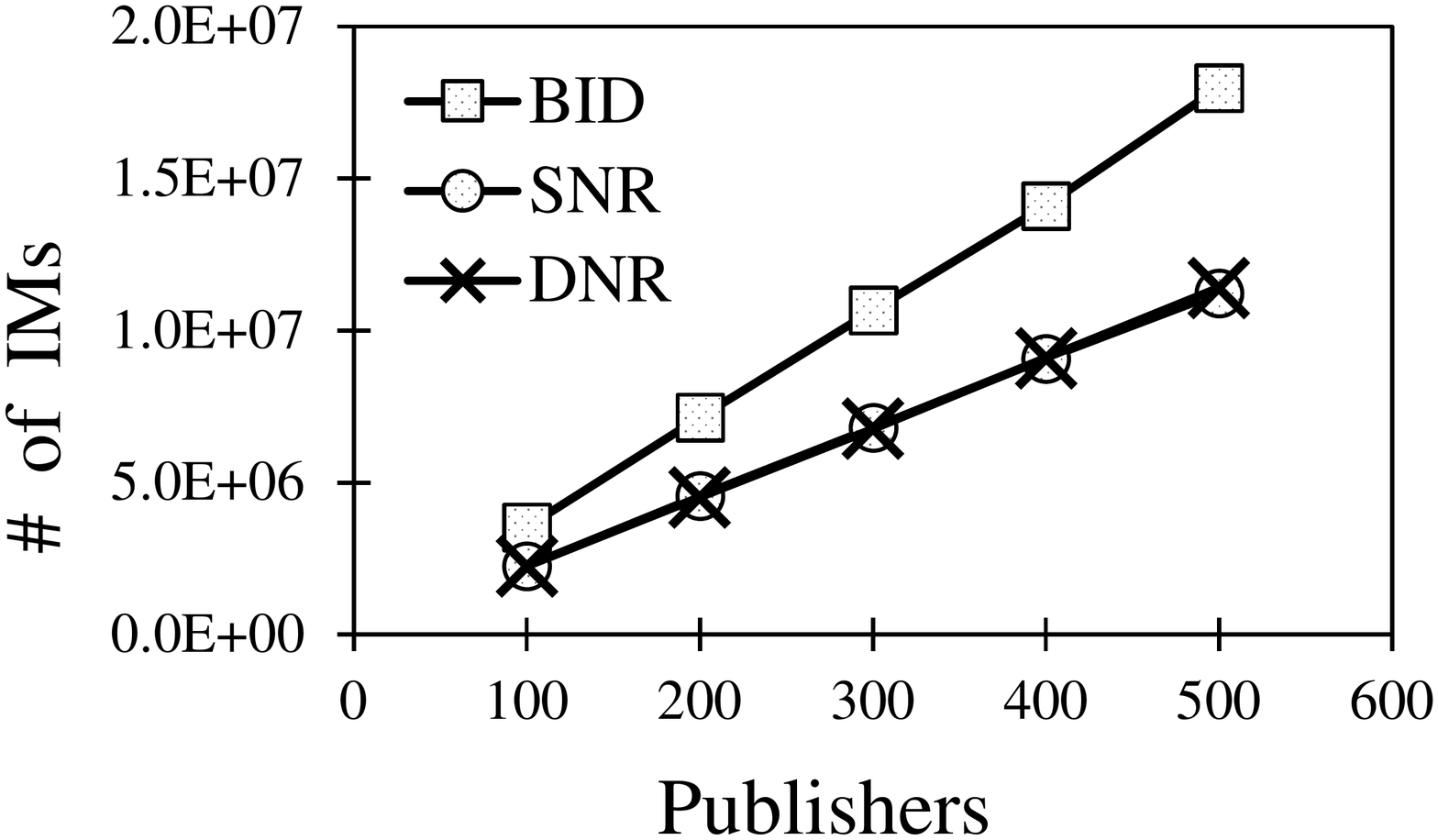}
		\caption{\small \# of IMs in notification routing}
		\label{fig:P5}
	\end{subfigure}
	~ %add desired spacing between images, e. g. ~, \quad, \qquad, \hfill etc.
	\begin{subfigure}[b]{0.24\textwidth}
		\includegraphics[trim=1cm 0.5cm 0.5cm 1cm, width=0.8\textwidth]{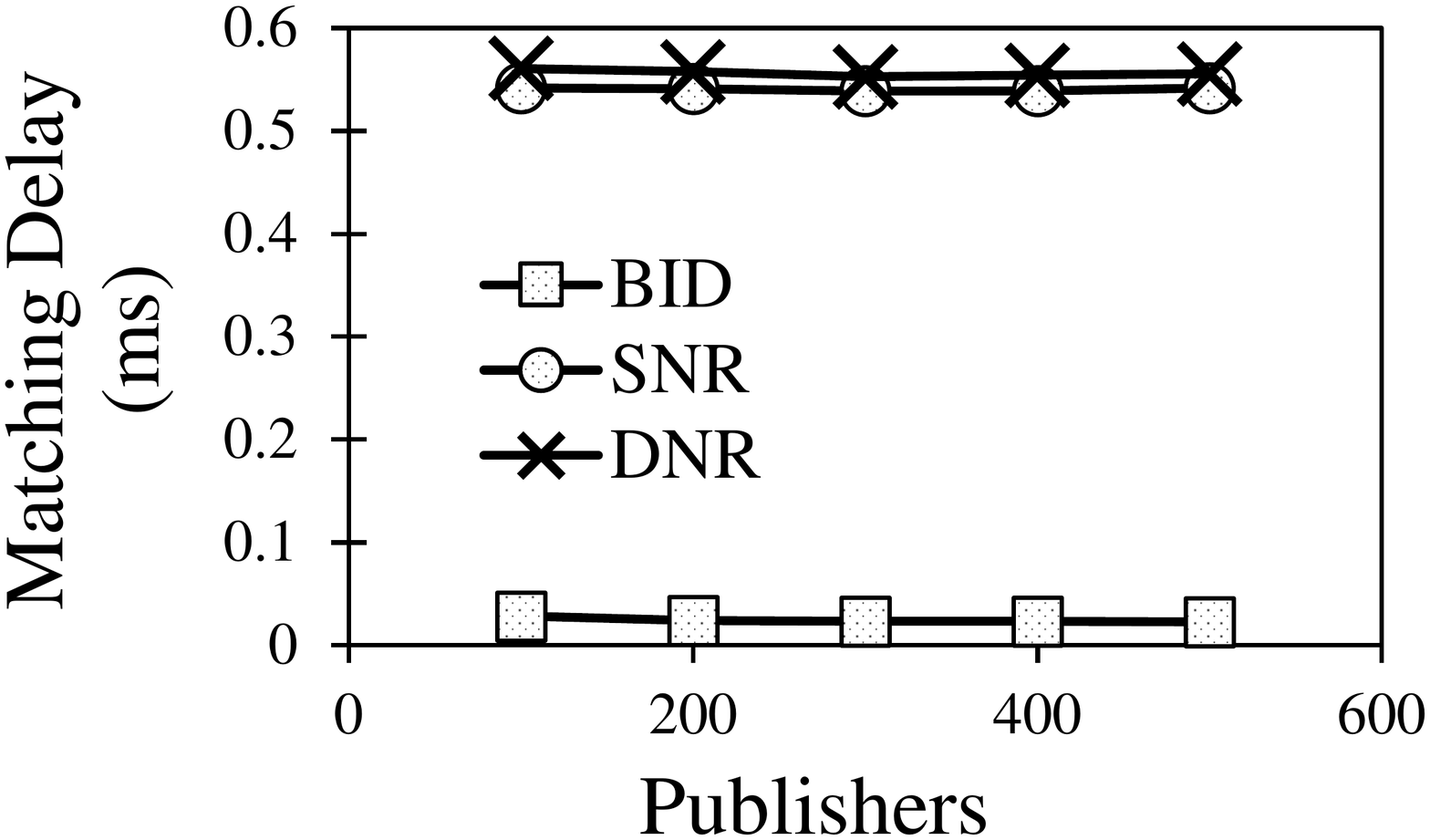}
		\caption{\small Matching delay}
		\label{fig:PN}
	\end{subfigure}
	\caption{(a) Matching delay in Subscriber Scalability. (b),(c),(d) Publisher Scalability in an unclustered and cluster-based SCOT.}
	\label{fig:publicationNumber}
\end{figure*}
\begin{figure*}
	\centering
	\begin{subfigure}[b]{0.3\textwidth}
		\includegraphics[trim=3cm 6.5cm 2cm 7cm, width=\textwidth]{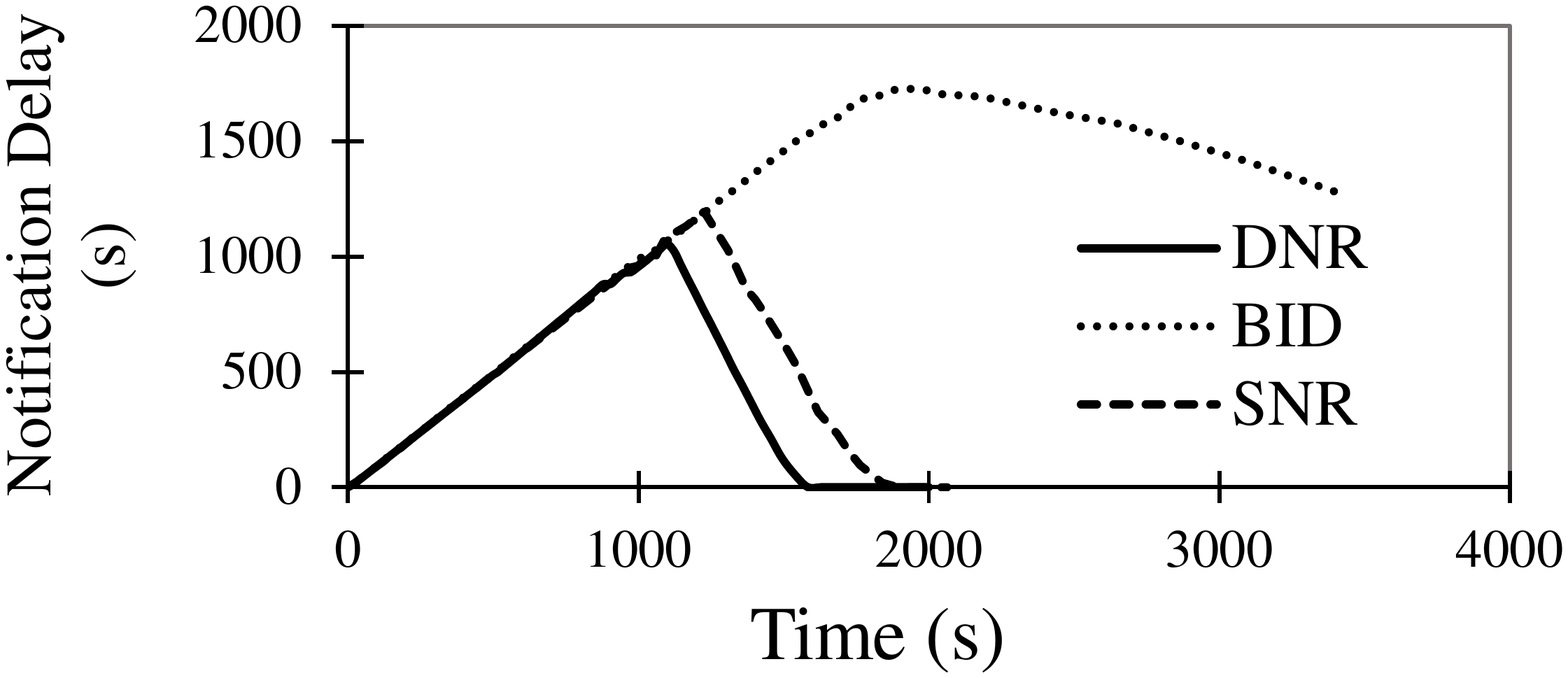}
		\caption{\small 100K messages per minute}
		\label{fig:P14}
	\end{subfigure}
	~
	\begin{subfigure}[b]{0.3\textwidth}
		\includegraphics[trim=3cm 6.5cm 2cm 7cm, width=\textwidth]{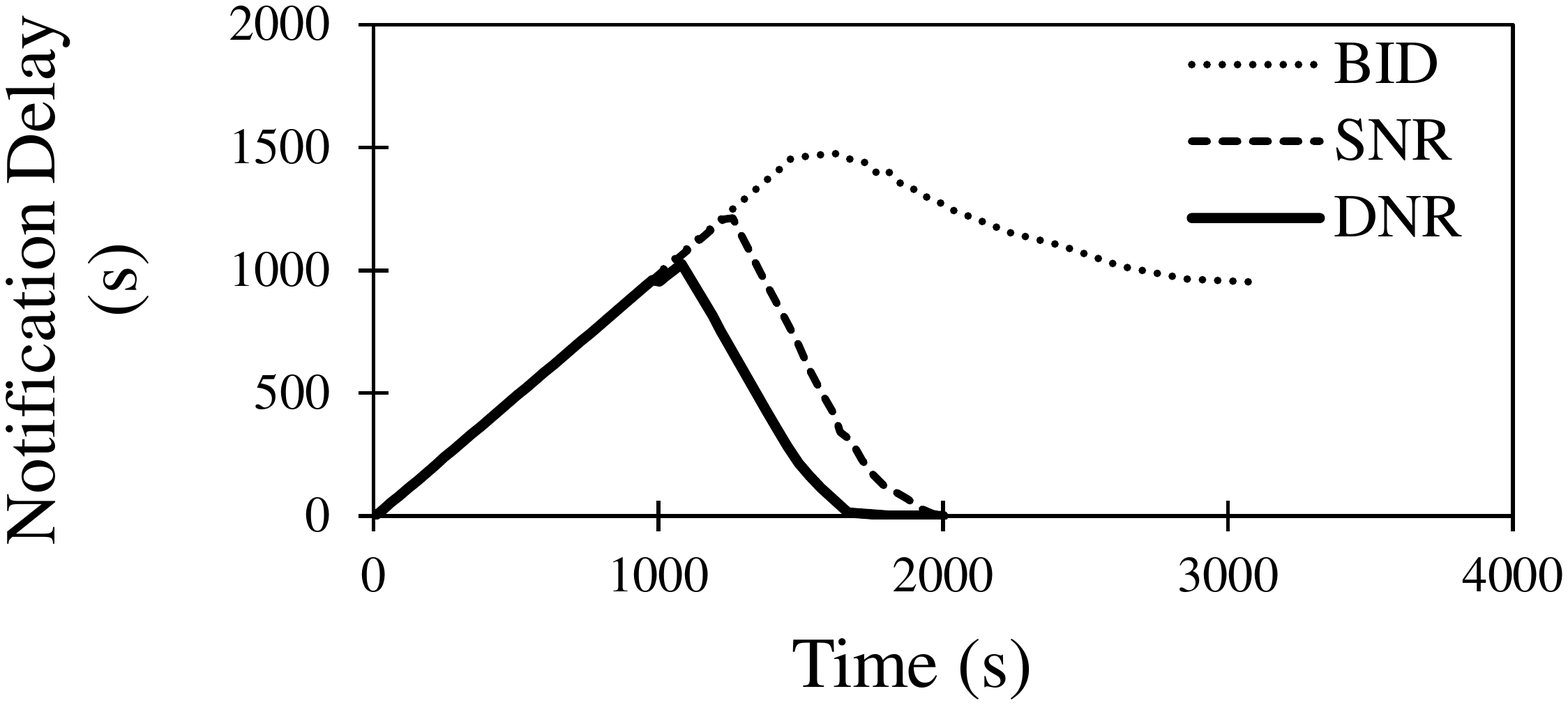}
		\caption{\small 80K messages per minute}
		\label{fig:P15}
	\end{subfigure}
	~
	\begin{subfigure}[b]{0.3\textwidth}
		\includegraphics[trim=3cm 6.5cm 2cm 7cm, width=\textwidth]{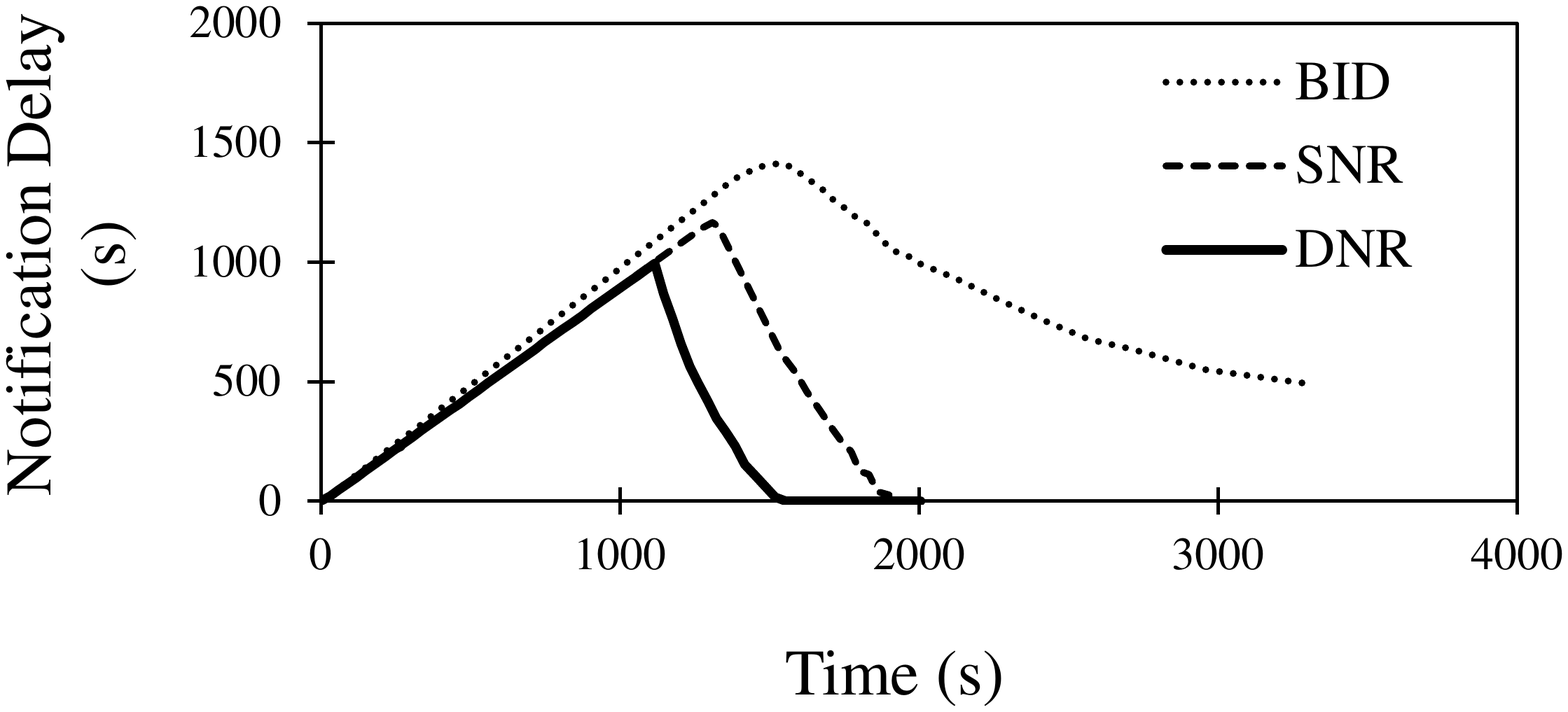}
		\caption{\small 60K messages per minute}
		\label{fig:P7}
	\end{subfigure}
	\caption{Stability analysis.}
	\label{fig:burst}
\end{figure*}
\subsection{Results}
The results presented in this section cover three important aspects of the evaluation: (i) \textit{Subscriber Scalability}, which studies the behaviour of the routing algorithms when the number of subscribers increases, while the number of publishers remains constant; (ii) \textit{Publisher Scalability} is about study of the algorithms with a varying number of publishers and constant number of subscribers; (iii) \textit{Burst Scenario}, in which an HRP starts sending notifications at a high rate and causes congestion in the output queues.\\
\textit{\textbf{Subscriber Scalability:}}
We gradually increased the number of subscribers from 500 to 10000, and used 100 publishers, each sending 1000 notifications at the rate of 60 notifications per minute (npm). All publishers start sending the notifications in the first 5 seconds after all the subscriptions register. We adopted this pattern of generating controlled workload to count IMs and notifications received by the subscribers. Fig. 9(a) shows that the number of IMs generated by SBP in clustered $\mathbb{S}_{e}$ (legends SNR, and DNR) are 89\% less than in unclustered $\mathbb{S}_{e}$ (legend BID). There are two reasons for this significant difference: (i) larger average lengths of subscription--trees, and (ii) extra IMs generated to discard duplicate subscriptions in unclustered $\mathbb{S}_{e}$. The average length of the subscription--trees generated by \texttt{OctopiS} is 14\% less than the PADRES. Furthermore, almost 80\% of the generated IMs are used to detect and discard duplicates in unclustered $\mathbb{S}_{e}$ \cite{Li_ADAP}. \texttt{OctopiS} uses a pattern of subscription broadcast that does not generate duplicates and extra IMs. Fig. 9(b) shows that the average subscription delay in clustered $\mathbb{S}_{e}$ is 77\% less than the unclustered $\mathbb{S}_{e}$. The difference is due to the larger lengths of subscription--trees and extra IMs generated in unclustered $\mathbb{S}_{e}$ to detect duplicate subscriptions. Importantly, there is no difference in subscription delay when SNR and DNR algorithms are used, since SBP does not use inter-cluster dynamic routing even if some output queues are congested. This approach generates subscription--trees of shortest lengths even if some links are overloaded. The average subscription delay in unclustered and clustered $\mathbb{S}_{e}$ is nearly constant. Fig. 9(c) shows that the end--to--end notification delay in SNR and DNR is less than BID--based routing approach. More specifically, SNR and DNR algorithms reduce the notification delay by 47\% when compared with BID--based routing algorithm. There are three reasons for this difference: (i) the larger length of subscription--trees in BID--based routing, (ii) high payload due to carrying BIDs of the matching subscriptions, and (iii) extra processing by brokers to prepare and split lists of BIDs down the routing paths to find the next destinations. In large networks with thousands of brokers in multiple data centres (e.g., in \cite{WSP_BING}), a notification may have to carry thousands of BIDs to identify routing paths. \texttt{OctopiS} does not require carrying BIDs with notifications (a lightweight bit vector $CBV_{p}$ is added when the notification routing is dynamic). Delay in static and dynamic routing is almost the same as this experiment does not generate dynamic routing paths for small workloads. Fig. 9(d) shows that the number of IMs generated by the same number of notifications (1000 per publisher) in SNR and DNR algorithms is 12\% less than BID--based approach. Again, the difference in the number of IMs is due to larger lengths of subscription--trees generated by the BID-based subscription routing. The difference between the number of generated IMs decreases with the increase in the number of subscribers because of the possibility of having less \textit{forwarder--only} brokers in a routing path. Fig. 10(a) shows that the average matching delay in BID--based routing is 25\% less than SNR and DNR algorithms, and increase almost linearly. In BID--based notification routing, only the host brokers of publishers and interested subscribers execute the matching process, while no matching occurs at the intermediate brokers \cite{Li_ADAP}, while SNR and DNR algorithms performs matching at each broker, which results in larger matching delays. The difference between matching delays in the three algorithms decrease with increase in the number of subscribers, which decreases forwarder--only brokers.\\
\textit{\textbf{Publisher Scalability:}}
We increase the number of publishers from 100 to 500 (each sending 500 notifications at a fixed rate of 100 npm) with 3000 subscribers. Fig. 10(b) shows that the average notification delay in SNR and DNR algorithms is 42\% less than BID--based routing algorithm. The difference is due to the larger length of subscription--trees and higher payload due to carrying BIDs with notifications. Because of the difference in lengths of subscription--trees, the number of IMs generated by BID--based routing is 36.4\% higher than SNR and DNR algorithms (Fig. 10(c)). The average matching delay in SNR and DNR algorithms is 265\% higher than BID--based algorithm (Fig. 10(d)). BID--based routing matches a notification only at the host brokers of publishers and interested subscribers, while SNR and DNR algorithms match a notification at each broker of a routing path. As the number of subscribers is constant, the matching delays introduced by three algorithms is linear.\\
\textit{\textbf{Stability Analysis:}}
The stability analysis tells how quickly a CPS system converges to a normal state after an HRP finishes sending notifications. To study this behaviour, we set the value of $\tau$ to 10 and $t_{w}$ to 50 milliseconds. We used 5000 subscribers, and 100 publishers each issued 2000 notifications at the rate of 60 npm. 0.2\% of the subscribers subscribe to receive notifications from the HRP. We execute three simulations with the HRP sending 100K notifications at rates of 100K, 80K, and 60K npm. All clusters of $\mathbb{S}_{e}$ are target of the HRP. Furthermore, the HRP and interested subscribers are hosted by different brokers exerting more load on iCOLs and aCOLs. The burst continued for 60 to 100 seconds depending on the rate of the HRP. Each point in the graphs in Fig. 11 is a maximum delivery delay of 1000 notifications received in a sequence. This approach helps in analysing the stability without graphing too many points. Each simulation is run until all notifications are received. Fig. 11 shows that DNR algorithm stabilized \texttt{OctopiS} before SNR algorithm, while BID--based routing algorithm is not able to stabilize PADRES for the same workload. On average, in the first 18 minutes and 30 seconds, the maximum delay of a notification (out of 1000) is the same in the three routing algorithms and no tendency toward stability is observed. This indicates that, due to the high rates of notifications, the state of the system (\texttt{OctopiS}) does not converge to normal until the condition $CE < 1$ is maintained for some time (on average, 16 minutes and 40 seconds for the three simulations). DNR starts stabilizing \texttt{OctopiS} before the other two algorithms. The average value of $Q_{\ell}$ of target links at the brokers in routing paths of notifications from the HRP when DNR is used is 48\% less than SNR and 59\% less than BID--based routing algorithms. There are 5 target clusters of the HRP and DNR tends to add the minimum number of copies of a notification when the output queues of the target links are congested. This decreased the length of $Q_{\ell}$ when compared with SNR and BID--based algorithms. The average notification delay in DNR algorithm when the notification rate is 100K is 13\%, and 58\% less than SNR, and BID--based routing algorithms, respectively. Similar improvements, when the rates are 80K, 60K npm, are 12.1\% and 53\%, and 11\% and 49.2\%, respectively. On average, for the three simulations, DNR algorithm stabilizes the system 239 seconds before SNR algorithm and generates only 0.32\% IMs more than SNR and 17.2\% less than BID--based routing algorithms. This shows that our approach of adding at most one copy of a notification when the output queues are congested significantly reduces delivery delay and queue length. Analysis of the collected data indicate that the number of notifications that have delivery delays less than 1 second in DNR, SNR and BID--based routing algorithms are 44.2\%, 39\% and 28\%, respectively. We also conducted several experiments with HRPs in each cluster and sending notifications with different rates. The performance difference between DNR and SNR algorithms decreases with an increase in the number of HRPs that start sending notifications simultaneously. This indicates that the improvement due to inter-cluster dynamic routing in \texttt{OctopiS} diminishes when more congested output queues and overloaded iCOLs appear. 
\section{Related Work}
Content-based routing in distributed broker-based CPS systems has been focus of many research efforts. SIENA \cite{SIENA_WIDE_AREA}, JEDI \cite{JEDI}, Rebeca \cite{Rebeca}, PADRES \cite{PADRESBookChapte}, Kyra \cite{Kyra}, and MEDYN \cite{MEDYN} are just few examples. Notifications routing in cyclic overlays has got a little attention from the research community. SIENA introduces a notifications routing scheme for cyclic overlays and detects duplicates using BIDs. Latency-based distance-vector algorithms generate routing paths for advertisement- and subscription-based CPS systems \cite{carz_thesis}. However, the algorithms do not generate subscription--trees of shortest lengths and, despite multi-path overlay networks, dynamic routing is not supported. Subscription--trees generated for CPS systems in \cite{carz_thesis} can be improved by periodically sending subscription messages to find links with the minimum latency; however, this refinement generates extra traffic in the network and is infeasible for a large network settings. Li et al \cite{Li_ADAP} offers dynamic routing without making updates in routing tables in advertisement-based publish/subscribe systems. A large number of IMs are generated in the advertisement broadcast process to detect duplicates. Unique path identifications are added in notifications for routing to interested subscribers. Dynamic routing relies on brokers with intersecting routing paths generated by multiple advertisements matching one subscription. Therefore dynamic routing for a subscription (or subscriber) matching with one advertisement is not possible. Furthermore, intersecting routing paths are not always possible even if a subscription matches with multiple advertisements \cite{Li_ADAP}. The probability of having brokers that publish intersecting routing paths also depends upon the number of multiple advertisements matching a subscription. Finally, a subscription has to be delivered multiple times to the broker that publishes more than one advertisement intersecting that subscription \cite{OctopiA_TR}. Shuping et al \cite{MERC} propose an approach which divides an overlay into interconnected clusters to apply content-based and destination based intra- and inter-cluster routing. Algorithms for routing using shortest paths are developed, however, the approach requires embedding routing path identifications, which increases payload and consume network bandwidth inefficiently. A broker has to be aware of all other brokers in its host cluster, which increases topology maintenance cost. Dynamic routing requires updates in routing tables to generate alternative routing paths, which increases network traffic and delivery delays.\\ 
\texttt{OctopiS} offers inter-cluster dynamic routing without requiring updates in routing tables. Thanks to the structuredness of clustered SCOT overlays, which provide multiple inter-cluster routing paths. \texttt{OctopiS} neither generates redundant IMs or duplicates in SBP nor requires BIDs. Notifications are delivered using subscription--trees of shortest lengths and without requiring unique identifications to identify routing paths to prevent loops. Brokers in SCOT overlay are decoupled and aware of their direct neighbours only, which requires a very low topology maintenance cost. All these properties make \texttt{OctopiS} scalable and suitable for large content-based routing networks.
\section{Conclusion}
In this paper, we present the design and evaluation of the first (subscription--based) CPS system, \texttt{OctopiS}, that offers inter--cluster dynamic routing of notifications without requiring updates in routing tables. The system uses a novel structured cyclic topology SCOT, which is constructed applying optimizations on Cartesian product of two graphs. A homogeneous clustering technique is introduced, which divides a SCOT into similar identifiable blocks of brokers. We further exploit the structuredness of clustered SCOT to generate subscription--trees of shortest lengths, without generating extra IMs to detect and discard duplicates. A static routing algorithm SNR uses subscription--trees of shortest lengths to send notifications to interested subscribers, while DNR algorithm offers inter--cluster dynamic routing using a lightweight bit vector mechanism. Both algorithms do not require a global knowledge of an overlay topology and dynamic routing is activated when congestion is detected in the output queues of the target links. The evaluation of SNR and DNR algorithms, and comparison with BID--based routing algorithm indicates that \texttt{OctopiS} scales better than state--of--the--art and suitable for large network settings.\\
\bibliographystyle{plain}
\bibliography{refs_p1}
\end{document}